\documentclass[twocolumn]{aastex631}
\usepackage{subfigure}
\usepackage{ulem}

\shorttitle{A Common Origin Between SNe Ibn and Some Fast Transients}
\shortauthors{Pellegrino et al.}
\graphicspath{{./}{figures/}}

\begin{document}

\title{Circumstellar Interaction Powers the Light Curves of Luminous Rapidly Evolving Optical Transients}

\correspondingauthor{Craig Pellegrino}
\email{cpellegrino@lco.global}

\author[0000-0002-7472-1279]{C. Pellegrino}
\affil{Las Cumbres Observatory, 6740 Cortona Drive, Suite 102, Goleta, CA 93117-5575, USA}
\affil{Department of Physics, University of California, Santa Barbara, CA 93106-9530, USA}

\author[0000-0003-4253-656X]{D. A. Howell}
\affil{Las Cumbres Observatory, 6740 Cortona Drive, Suite 102, Goleta, CA 93117-5575, USA}
\affil{Department of Physics, University of California, Santa Barbara, CA 93106-9530, USA}

\author[0000-0001-8764-7832]{J. Vink\'o}
\affil{ Konkoly Observatory,  CSFK, Konkoly-Thege M. \'ut 15-17, Budapest, 1121, Hungary}
\affil{ELTE E\"otv\"os Lor\'and University, Institute of Physics, P\'azm\'any P\'eter s\'et\'any 1/A, Budapest, 1117 Hungary}
\affil{Department of Optics \& Quantum Electronics, University of Szeged, D\'om t\'er 9, Szeged, 6720, Hungary}
\affil{Department of Astronomy, University of Texas at Austin, 2515 Speedway, Stop C1400, Austin, Texas 78712-1205, USA}

\author[0000-0002-3884-5637]{A. Gangopadhyay}
\affil{Hiroshima Astrophysical Science Centre, Hiroshima University, Kagamiyama, Higashi-Hiroshima, Hiroshima 739-8526, Japan}

\author[0000-0002-1089-1519]{D. Xiang}
\affil{Physics Department and Tsinghua Center for Astrophysics (THCA), Tsinghua University, Beijing, 100084, China}

\author[0000-0001-7090-4898]{I. Arcavi}
\affil{School of Physics and Astronomy, Tel Aviv University, Tel Aviv 69978, Israel}
\affil{CIFAR Azrieli Global Scholars program, CIFAR, Toronto, Canada}

\author[0000-0001-6272-5507]{P. Brown}
\affil{Department of Physics and Astronomy, Texas A\&M University, 4242 TAMU, College Station, TX 77843, USA}
\affil{George P. and Cynthia Woods Mitchell Institute for Fundamental Physics \& Astronomy, USA}

\author[0000-0003-0035-6659]{J. Burke}
\affil{Las Cumbres Observatory, 6740 Cortona Drive, Suite 102, Goleta, CA 93117-5575, USA}
\affil{Department of Physics, University of California, Santa Barbara, CA 93106-9530, USA}

\author[0000-0002-1125-9187]{D. Hiramatsu}
\affil{Las Cumbres Observatory, 6740 Cortona Drive, Suite 102, Goleta, CA 93117-5575, USA}
\affil{Department of Physics, University of California, Santa Barbara, CA 93106-9530, USA}

\author[0000-0002-0832-2974]{G. Hosseinzadeh}
\affil{Center for Astrophysics, Harvard \& Smithsonian, 60 Garden Street, Cambridge, MA 02138-1516, USA}

\author{Z. Li}
\affil{Key Laboratory of Optical Astronomy, National Astronomical Observatories, Chinese Academy of Sciences, Beijing 100101, China}

\author[0000-0001-5807-7893]{C. McCully}
\affil{Las Cumbres Observatory, 6740 Cortona Drive, Suite 102, Goleta, CA 93117-5575, USA}
\affil{Department of Physics, University of California, Santa Barbara, CA 93106-9530, USA}

\author[0000-0003-1637-267X]{K. Misra}
\affil{Aryabhatta Research Institute of Observational Sciences, Manora Peak, Nainital 263 002 India}

\author{M. Newsome}
\affil{Las Cumbres Observatory, 6740 Cortona Drive, Suite 102, Goleta, CA 93117-5575, USA}
\affil{Department of Physics, University of California, Santa Barbara, CA 93106-9530, USA}

\author{E. Padilla Gonzalez}
\affil{Las Cumbres Observatory, 6740 Cortona Drive, Suite 102, Goleta, CA 93117-5575, USA}
\affil{Department of Physics, University of California, Santa Barbara, CA 93106-9530, USA}

\author[0000-0001-9227-8349]{T. A. Pritchard}
\affil{Center for Cosmology and Particle Physics, Department of Physics, New York University, 726 Broadway, New York, NY 10003, USA}

\author[0000-0001-8818-0795]{S. Valenti}
\affil{Department of Physics, University of California, Davis, CA 95616, USA}

\author[0000-0002-7334-2357]{X. Wang}
\affil{Physics Department and Tsinghua Center for Astrophysics (THCA), Tsinghua University, Beijing, 100084, China}
\affil{Beijing Planetarium, Beijing Academy of Sciences and Technology, Beijing, 100044, China}

\author{T. Zhang}
\affil{Key Laboratory of Optical Astronomy, National Astronomical Observatories, Chinese Academy of Sciences, Beijing 100101, China}

\begin{abstract}

Rapidly evolving transients, or objects that rise and fade in brightness on timescales two to three times shorter than those of typical Type Ia or Type II supernovae (SNe), have uncertain progenitor systems and powering mechanisms. Recent studies have noted similarities between rapidly evolving transients and Type Ibn SNe, which are powered by ejecta interacting with He-rich circumstellar material (CSM). In this work we present multiband photometric and spectroscopic observations from Las Cumbres Observatory and Swift of four fast-evolving Type Ibn SNe. We compare these observations with those of rapidly evolving transients identified in the literature. We discuss several common characteristics between these two samples, including their light curve and color evolution as well as their spectral features. To investigate a common powering mechanism we construct a grid of analytical model light curves with luminosity inputs from CSM interaction as well as $^{56}$Ni radioactive decay. We find that models with ejecta masses of $\approx 1-3$ M$_\odot$, CSM masses of $\approx 0.2-1$ M$_\odot$, and CSM radii of $\approx 20-65$ au can explain the diversity of peak luminosities, rise times, and decline rates observed in Type Ibn SNe and rapidly evolving transients. This suggests that a common progenitor system \textemdash the core collapse of a high-mass star within a dense CSM shell \textemdash can reproduce the light curves of even the most luminous and fast-evolving objects, such as AT\,2018cow. This work is one of the first to reproduce the light curves of both SNe Ibn and other rapidly evolving transients with a single model.

\end{abstract}

\keywords{Supernovae (1668) --- Core-collapse supernovae (304) --- Circumstellar matter (241)}

\section{Introduction}

Over the last several years time-domain surveys, including Panoramic Survey Telescope and Rapid Response System, Pan-STARRS1 \citep[PS1;][]{Kaiser2010}, the Dark Energy Survey \citep[DES;][]{Flaugher2005}, and the Zwicky Transient Facility \citep[ZTF;][]{Bellm2019,Graham2019}, have led to the discovery of thousands of astronomical transients. Among these discoveries have been objects that are more luminous and evolve more rapidly than other known classes of supernovae (SNe). Samples of these rapidly evolving (hereafter ``fast") transients have been identified in PS1 \citep[hereafter D14]{Drout2014}, the Supernova Legacy Survey \citep{Arcavi2016}, DES \citep{Pursiainen2018}, the Kepler mission \citep{Rest2018}, and ZTF \citep{Ho2021}, among others. Although their exact classification has varied, broadly they display rises to peak brightness in fewer than five days and declines from peak to half-peak brightness in fewer than ten days. D14 was one of the first to identify a large sample of fast transients that had a time above half their maximum brightness, t$_{1/2}$, of $\lesssim$\,12 days and absolute magnitude -16.5 $\lesssim$ M $\lesssim$ -20.

Most of the fast transients identified to date have been found at cosmological distances, i.e. $d_L$  $\gtrsim$ 200 Mpc, making full multiband studies of these objects difficult. This changed with the discovery of AT\,2018cow, a fast transient identified at a redshift $z=0.014$ \citep{Benetti2018}, or luminosity distance $d_L = 60 $ Mpc \citep{Prentice2018}. AT\,2018cow presented the first opportunity for a true multiband study, from radio to $\gamma$ ray, of a nearby fast transient. Observations of strong X-ray emission \citep{Margutti2019}, an initially hot and featureless spectrum \citep{Prentice2018}, and a receding photosphere \citep{Perley2019} all affected the physical interpretation of the progenitor system of this fast transient. Since the discovery of AT\,2018cow other similar transients have been discovered at higher redshifts \citep{Coppejans2020,Ho2020,Perley2021}.

Due to their high peak luminosities and rapid evolution, modeling the powering mechanism of these fast transients has proven difficult. A $^{56}$Ni-decay powering source is impossible to reconcile with both the peak luminosities and rapid evolution of these objects. Other possible powering sources include the thermonuclear explosion of a white dwarf within an H-rich envelope \citep{Arcavi2016}, magnetar spin-down \citep{Prentice2018}, a tidal disruption event by an intermediate-mass black hole \citep{Perley2019}, and interaction with circumstellar material \citep[CSM;][]{Drout2014,RiveraSandoval2018}. These various models each have their advantages and drawbacks when compared to the complex temporal evolution of AT\,2018cow.

More recently, similar characteristics have been noticed between certain fast transients, specifically AT\,2018cow, and Type Ibn supernovae \citep[SNe Ibn; ][]{Fox2019,Xiang2021}. SNe Ibn are rare but well studied \citep[hereafter H17]{Pastorello2008,Hosseinzadeh2017}. There is evidence that their progenitor systems are high-mass stars, such as Wolf-Rayet (WR) stars, that undergo significant mass loss in a short period of time before explosion \citep[but see also \citealt{Hosseinzadeh2019}]{Smith2006,Foley2007,Smith2012}. In many cases their early-time spectra show hot blue continua superimposed with emission lines of He I and He II, indicating interaction with a CSM composed of material possibly stripped from a massive star. In particular, SNe Ibn show similar rise times, peak luminosities, and decline times when compared with fast transients \citep{Fox2019,Clark2020,Xiang2021}, hinting that some SNe Ibn may be included in samples of photometrically identified fast transients.

In one of the largest samples of fast transients to date, \citet{Ho2021} published observations of 42 objects discovered during Phase I of ZTF, some of which were observed by the Bright Transient Survey \citep{Fremling2020,Perley2020}, with times above half the maximum brightness of fewer than 12 days. Of these objects, 20 were spectroscopically classified. The objects with spectra in the sample primarily consist of core-collapse SNe such as Type IIb SNe (SNe IIb) that are powered by shock-cooling emission at early times, interaction-powered objects mainly classified as SNe Ibn, and more extreme objects such as AT\,2018cow. This study\textemdash one of the first to present a large sample of spectroscopically classified fast-evolving transients\textemdash points toward fast transients being a heterogeneous class of objects, with many powered by CSM interaction. However, many other objects in this sample, including the most luminous and fastest-evolving transients similar to AT\,2018cow, remain spectroscopically unclassified, and their powering mechanisms are uncertain.

As time-domain surveys discover more SN candidates than can be spectroscopically classified, it is important to investigate whether a single progenitor system and powering mechanism can explain the observed properties of objects in different regions of the fast transient parameter space. In this work, we compare photometry and spectra between SNe Ibn and other photometrically classified fast transients in literature (i.e. t$_{1/2}$ $\lesssim$ 12 days) in order to explore a common progenitor system for these objects. We identify four fast-evolving SNe Ibn with Las Cumbres Observatory \citep[LCO;][]{Brown2013} observations and compare these SNe with a sample of fast transients from D14 and AT\,2018cow. As CSM interaction at early times and $^{56}$Ni decay at late times are the proposed powering sources of SNe Ibn, we investigate whether they can reproduce the light curves of other fast transients, as well. We model the bolometric luminosities of the objects in our sample with inputs from CSM interaction plus $^{56}$Ni decay. We calculate rise times, peak luminosities, and decline times for these models and compare them with light-curve parameters for SNe Ibn and other photometrically classified fast transients. 

This paper is organized as follows. In Section \ref{sec:sample} we discuss the objects and data in our sample of fast transients and SNe Ibn. In Section \ref{sec:data} we detail the photometric and spectroscopic analysis of these objects. We describe our model light curves and compare them to data in Section \ref{sec:csmmodels}. We discuss a possible common progenitor system between SNe Ibn and some fast transients in Section \ref{sec:discussion}. Finally, we conclude in Section \ref{sec:conclusions}.

\begin{figure*}
    \centering
    \subfigure[SN 2019deh]{\includegraphics[width=0.48\textwidth]{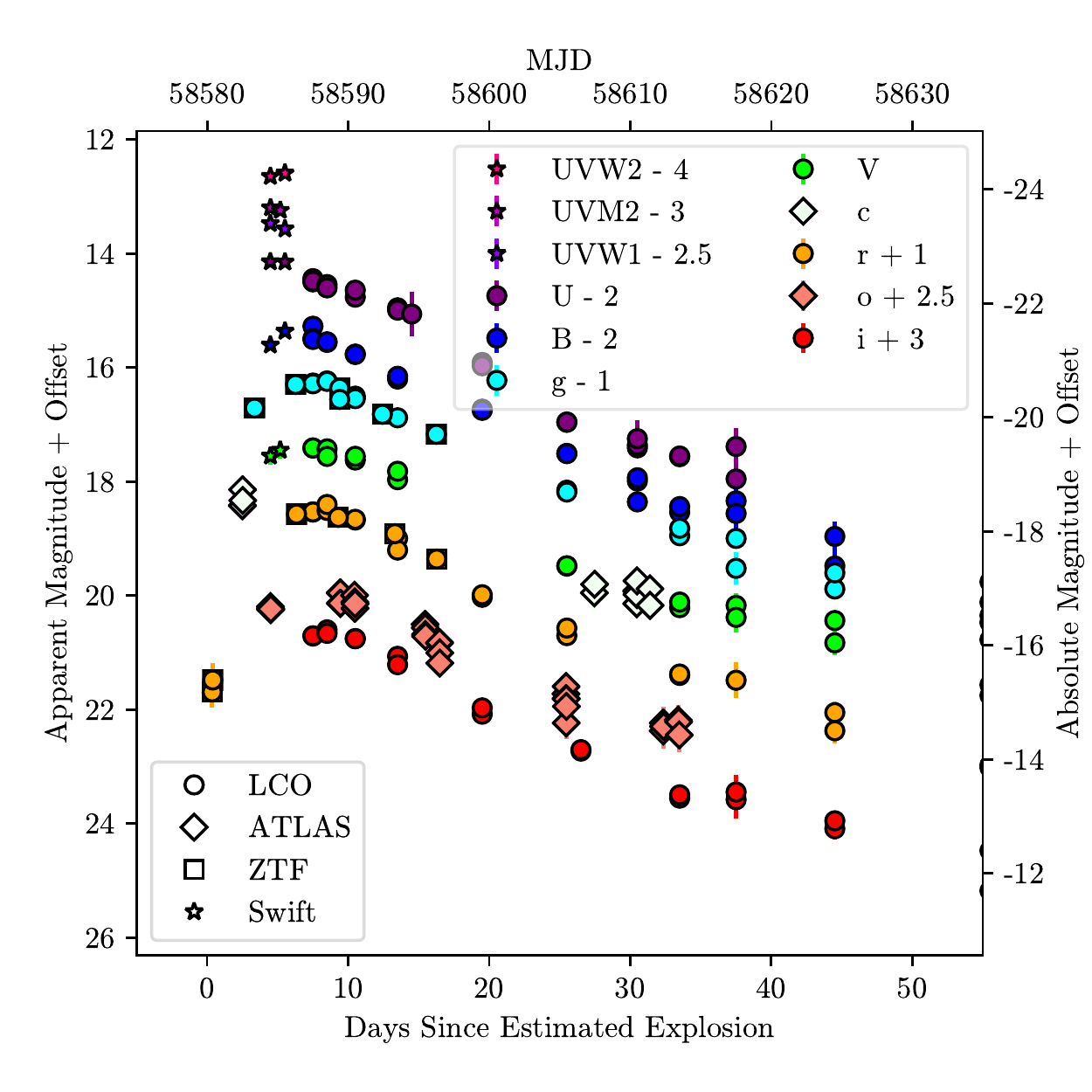}}\label{fig:sn2019dehlc}
    \subfigure[SN 2021jpk]{\includegraphics[width=0.48\textwidth]{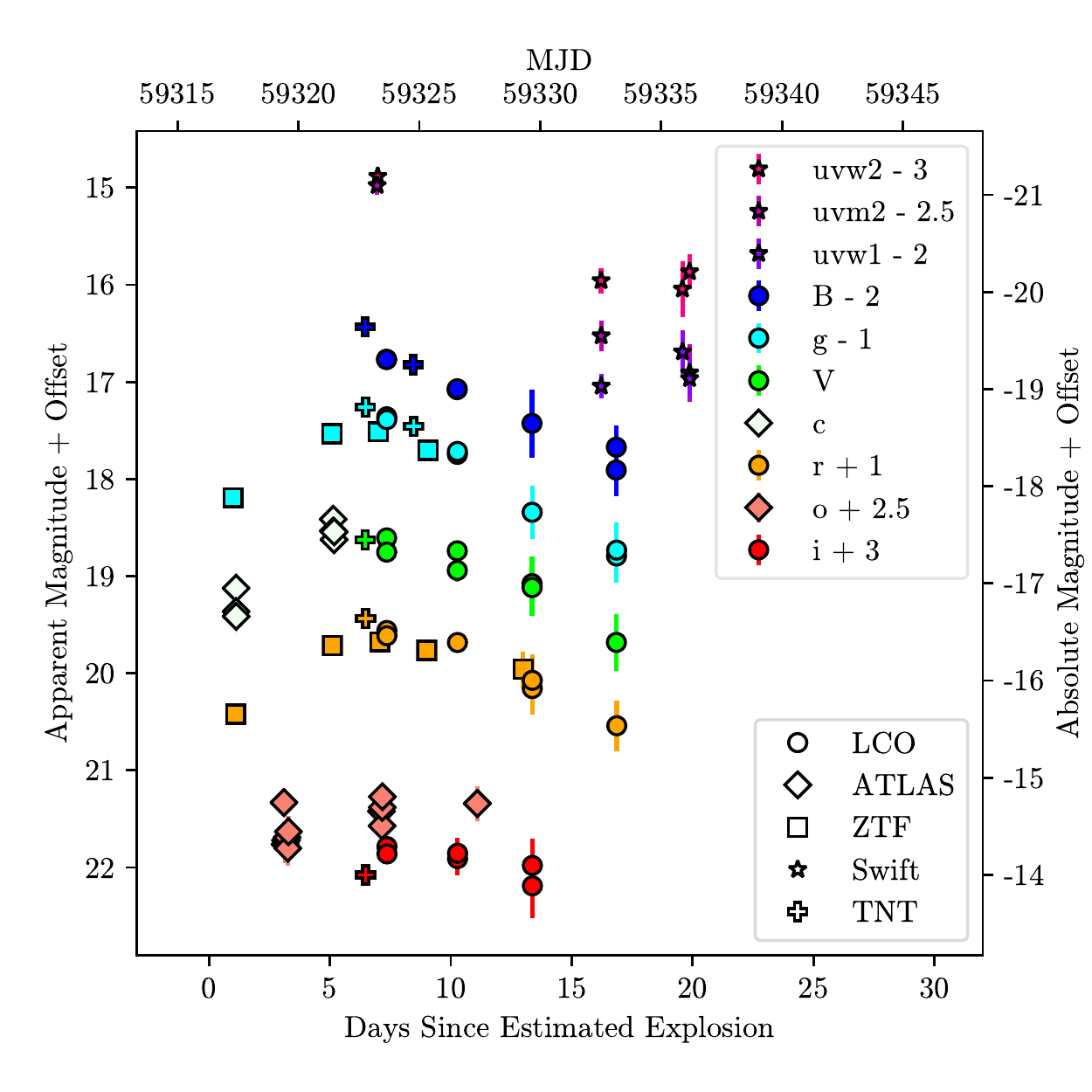}}\label{fig:sn2021jpklc}
    \caption{Observed UV and optical light curves of the Type Ibn (a) SN\,2019deh and (b) SN\,2021jpk from LCO (circles), ATLAS (diamonds), ZTF (squares), Swift (stars), and TNT (plus signs). All photometry have been corrected for MW extinction.}
    \label{fig:allbandlcs}
\end{figure*}

\section{Observations and Sample Description}\label{sec:sample}

Throughout this work we compare a sample of SNe Ibn observed by LCO with fast transients from D14 as well as AT\,2018cow. Details of each sample, including selection criteria for our SNe Ibn, are presented below.

\subsection{Fast-evolving SNe Ibn} \label{subsec:lcodata}

We begin by identifying four fast-evolving SNe Ibn observed by LCO through the Global Supernova Project (GSP). These objects were chosen because they all have optical and ultraviolet (UV) observations beginning at or before maximum light, spectra obtained within a few days of maximum light, and $\textit{g}$-band decline rates greater than 0.1 mag day$^{-1}$, which is the average decline rate for SNe Ibn \citep{Hosseinzadeh2017}. Because of their faster-than-average evolution, we classify these objects as fast-evolving SNe Ibn. Two (SN\,2019uo and SN\,2019wep) have LCO observations and data-reduction details presented in other works \citep[][Gangopadhyay 2021, in preparation]{Gangopadhyay2020}. In this work we present LCO photometry and spectra of SN\,2019deh and SN\,2021jpk, two additional fast-evolving SNe Ibn. ZTF observations of SN\,2019deh were discussed in \citet{Ho2021} while SN\,2021jpk has no published data thus far.

SN\,2019deh and SN\,2021jpk were discovered by ZTF on MJD 58580.36 (2019 April 7.36 UT) and MJD 59317.29 (2021 April 13.29) at \textit{r}-band magnitude 20.75 $\pm$ 0.28 and \textit{g}-band magnitude 19.26 $\pm$ 0.09, respectively. Assuming a standard cosmology with $H_0$ = 73 km s$^{-1}$ Mpc$^{-1}$, $\Omega_m =$ 0.27, and $\Omega_\Lambda =$ 0.73, the luminosity distances we use for SN\,2019deh and SN\,2021jpk are 237 Mpc and 164 Mpc, respectively \citep{Beers1995,AdelmanMcCarthy2007}. Due to the rapid evolution of these objects, the first LCO observations were not obtained until around the time of maximum brightness and continued for the next several weeks.

LCO light curves for both objects along with detections from ZTF, the Asteroid Terrestrial-impact Last Alert System \citep[ATLAS;][]{Tonry2018}, Swift, and the Tsinghua NAOC Telescope \citep[TNT;][]{Huang2012} are shown in Figure \ref{fig:allbandlcs}. We correct the photometry for Milky Way (MW) dust extinction assuming $A_V = 0.0772$ and $A_V = 0.0555$ from the dust maps of \citet{Schlafly2011} for SN\,2019deh and SN\,2021jpk, respectively. Due to the relatively large offset between SN\,2019deh and its host galaxy, we assume a negligible host-galaxy extinction. For SN\,2021jpk, we attempt to estimate the host-galaxy extinction by comparing its \textit{B - V} colors to those of Type Ibn SN\,2010al, which is spectroscopically similar to SN\,2021jpk \citep{Taubenberger2021}. After correcting for MW extinction the colors of both objects are consistent. Additionally, the spectrum of SN 2021jpk shows no host Na ID absorption; therefore we assume negligible host extinction. To estimate the time of explosion and time of maximum light for SN\,2019deh we fit a second-order polynomial to the ATLAS fluxes in \textit{o}-band during the first 15 days of observations. The estimated explosion time, $t_{exp}$, is MJD 58579.99 $\pm$ 0.25 and time of maximum brightness, $t_{peak}$, is MJD 58588.5 $\pm$ 0.65. Due to the sparse light-curve sampling at early times we take the average of the last nondetection and first ZTF detection of SN\,2021jpk as a conservative estimate of its explosion date, $t_{exp} = $ MJD 59316.3 $\pm$ 0.99. From fitting a second-order polynomial to the peak of the light curve we estimate $t_{peak} = \text{MJD } 59324.10 \pm 0.45$. The parameters for the objects in our SNe Ibn sample are summarized in Table \ref{tab:ibnsample}.

\subsubsection{Optical Photometry}
LCO \textit{UBgVri}-band data were obtained with the Sinistro camera on LCO 1m telescopes. Using the \texttt{lcogtsnpipe} photometric data-reduction pipeline \citep{Valenti2016} point-spread function (PSF) fitting was performed on LCO images to extract PSF magnitudes \citep{Stetson1987}. The \textit{UBV}-band photometry was calibrated to Vega magnitudes using Landolt standard fields \citep{Landolt1992}, while \textit{gri}-band photometry was calibrated to AB magnitudes using the Sloan Digital Sky Survey \citep{Smith2002}. Color terms for each epoch were computed using these standards. Background subtraction was performed on four of the objects (SN\,2019uo, SN\,2019wep, AT\,2018cow, and SN\,2021jpk) due to their proximity to their host galaxies. Template images were obtained after the SNe had faded and subtraction was performed using \texttt{PyZOGY} \citep{Guevel2017}, an implementation of the algorithm described in \cite{Zackay2016}.

We also obtained two epochs of \textit{BgVri}-band photometry of SN 2021jpk using the 0.8 m TNT. All images were processed using standard IRAF\footnote{IRAF is distributed by the National Optical Astronomy Observatories, which are operated by the Association of Universities for Research in Astronomy, Inc., under cooperative agreement with the National Science Foundation (NSF).} techniques. PSF photometry was calibrated to standard stars and converted to \textit{BgVri} magnitudes. Because the SN signal was strong at the time of observation, background subtraction was not performed. All optical photometry is presented in Table \ref{tab:optphot}.

\subsubsection{Swift Photometry}
We also present UV observations obtained with the Ultraviolet and Optical Telescope \citep[UVOT;][]{Roming2005} on the Niel Gehrels Swift Observatory \citep{Gehrels2004} for SN\,2021jpk. Images were obtained in the \textit{uvw2}, \textit{uvm2}, and \textit{uvw1} filters beginning MJD 59323.3, coincident with the time of maximum light. The data were reduced using the data-reduction pipeline of the Swift Ultraviolet/Optical Supernova Archive \citep{Brown2014} using the aperture corrections and zero-points of \citet{Breeveld2011}. Galaxy subtraction was not performed. Swift photometry in Vega magnitudes are presented in Table \ref{tab:uvphot}.

\subsubsection{Optical Spectra}
LCO spectra covering the optical range from 3500 to 10,000 $\text{\AA }$ at a resolution R $\approx$ 300-600 were obtained with the Folded Low Order whYte-pupil Double-dispersed Spectrograph (FLOYDS) spectrographs on the Faulkes Telescope North and Faulkes Telescope South through the GSP. Data were reduced using the \texttt{floydsspec} custom pipeline, which performs flux and wavelength calibration, cosmic-ray removal, and spectrum extraction\footnote{https://github.com/svalenti/FLOYDS\_pipeline/}. Details of the spectra shown in this work are presented in Table \ref{tab:spec} and a discussion of their features used for classification is given in Section \ref{subsec:spec}.

\subsection{D14 Fast Transients}

Throughout this work we compare our fast-evolving SNe Ibn sample to the gold and silver samples of rapidly evolving transients presented in D14 from the PS1 Median Deep Survey. These ten objects were confirmed to be extragalactic in origin and satisfied three criteria: (1) they rose $\gtrapprox 1.5$ mag in the previous nine days before maximum light; (2) they declined $\gtrapprox 1.5$ mag in the 25 days after maximum light; and (3) they appeared in at least three consecutive observations. These criteria were selected in order to exclude the most common SNe subtypes.

The objects in the gold and silver samples have $t_{1/2}$ $\lesssim$ 12 days, a median redshift of $z = 0.275$ and a median of 19 photometric detections across the optical region of the electromagnetic spectrum. Five have spectroscopic observations within several days of maximum light. All were observed in the \textit{g$_{P1}$r$_{P1}$i$_{P1}$z$_{P1}$} filters. Additional observations for several objects were obtained with the Gemini GMOS \citep{Hook2004} and Magellan IMACS \citep{Dressler2006} instruments in \textit{ri}-band. Data were reduced as described in D14. 

We correct photometry for MW extinction using the \textit{E(B-V)} values listed in Table 1 of D14. We also use the D14 luminosity distances when calculating the bolometric luminosities of these objects. Since PS1 does not observe in all filters every night, we follow the process described in D14 to interpolate photometric observations to a common epoch. Due to the rapid evolution of these objects, we only interpolate observations that were taken within a day of a \textit{g$_{P1}$}-band detection. For each common epoch with observations in at least three filters, we fit a blackbody spectral energy distribution (SED) to the rest-frame fluxes in order to calculate bolometric luminosities. The results from our best-fit blackbody SEDs are consistent with those presented in D14.

\subsection{AT 2018cow}\label{subsec:cow}

\begin{figure}
    \centering
    \includegraphics[width=0.48\textwidth]{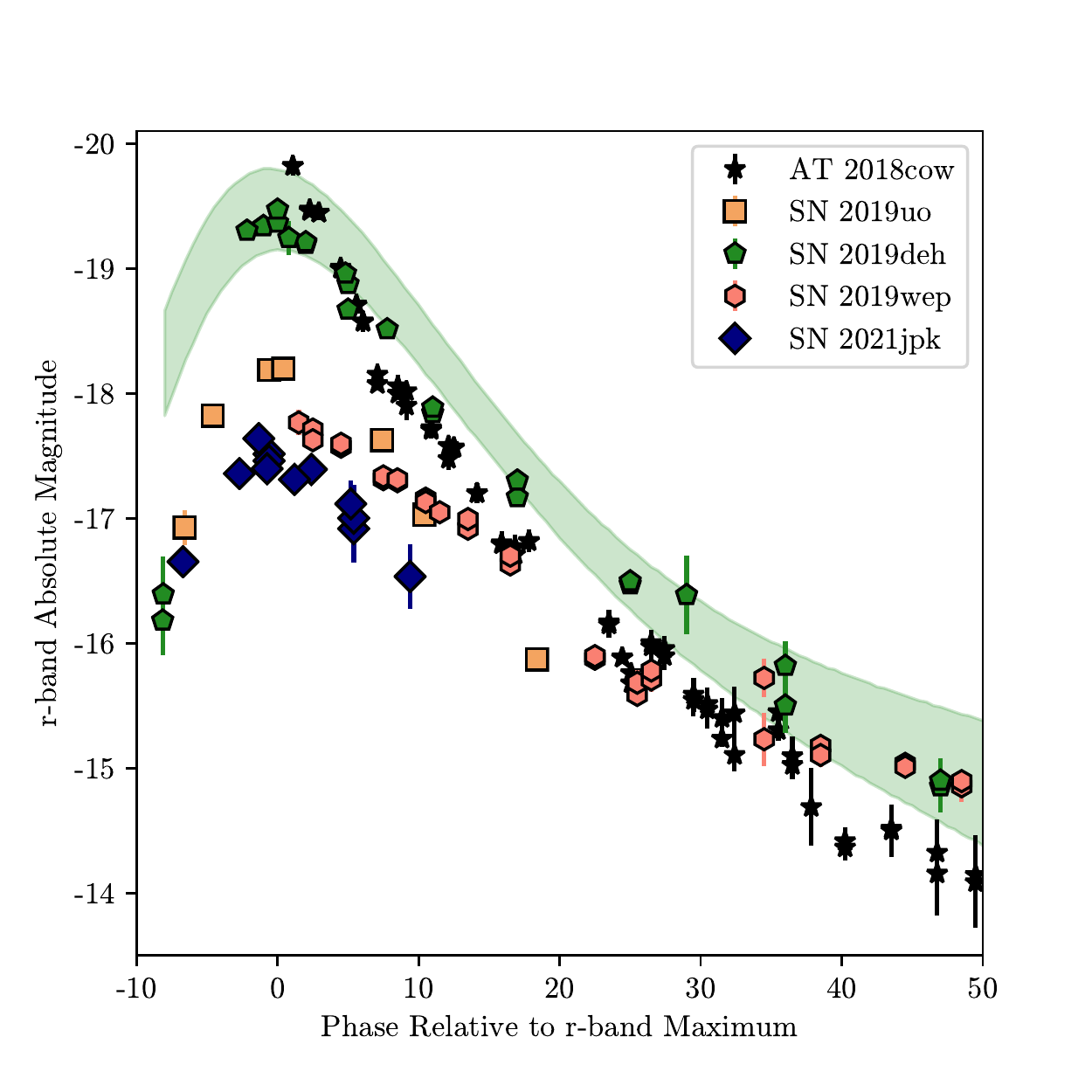}
    \caption{\textit{r}-band absolute magnitude light curves for the fast-evolving SNe Ibn as well as AT\,2018cow. All photometry have been corrected for MW extinction. The shaded region is the \textit{R}-band SNe Ibn template from H17. The objects presented here tend to evolve more rapidly, and have a wider range of peak luminosities, than the H17 template.}
    \label{fig:absmags}
\end{figure}

We also compare the SNe Ibn and D14 fast transients to LCO observations of AT\,2018cow. AT\,2018cow was discovered by ATLAS on MJD 58285.44 in CGCG 137-068 at a redshift of $z = 0.014145$ \citep{Prentice2018}. Due to its high luminosity and recent nondetection about four days prior, AT\,2018cow was quickly identified as an unusual transient \citep{Smartt2018}. Rapid follow-up across the electromagnetic spectrum began soon after discovery \citep{Prentice2018,RiveraSandoval2018,Margutti2019,Perley2019,Xiang2021}, making it the best-observed fast transient to date. 

LCO began observing AT\,2018cow on MJD 58288.07 with daily photometric and spectroscopic cadences. \textit{UBgVri}-band images and optical spectra were taken nearly continuously for the first two months of the object's evolution before it became too faint to observe. LCO data of AT\,2018cow, as well as a description of the data-reduction process, are presented in \citet{Xiang2021}. Throughout this paper we use the bolometric luminosities calculated in \citet{Xiang2021} as well as the rise time, decline time, and peak absolute magnitudes presented in \citet{Prentice2018} and \citet{Perley2019}.

\section{Data Analysis}\label{sec:data}

\begin{figure*}
    \centering
    \includegraphics[width=0.8\textwidth]{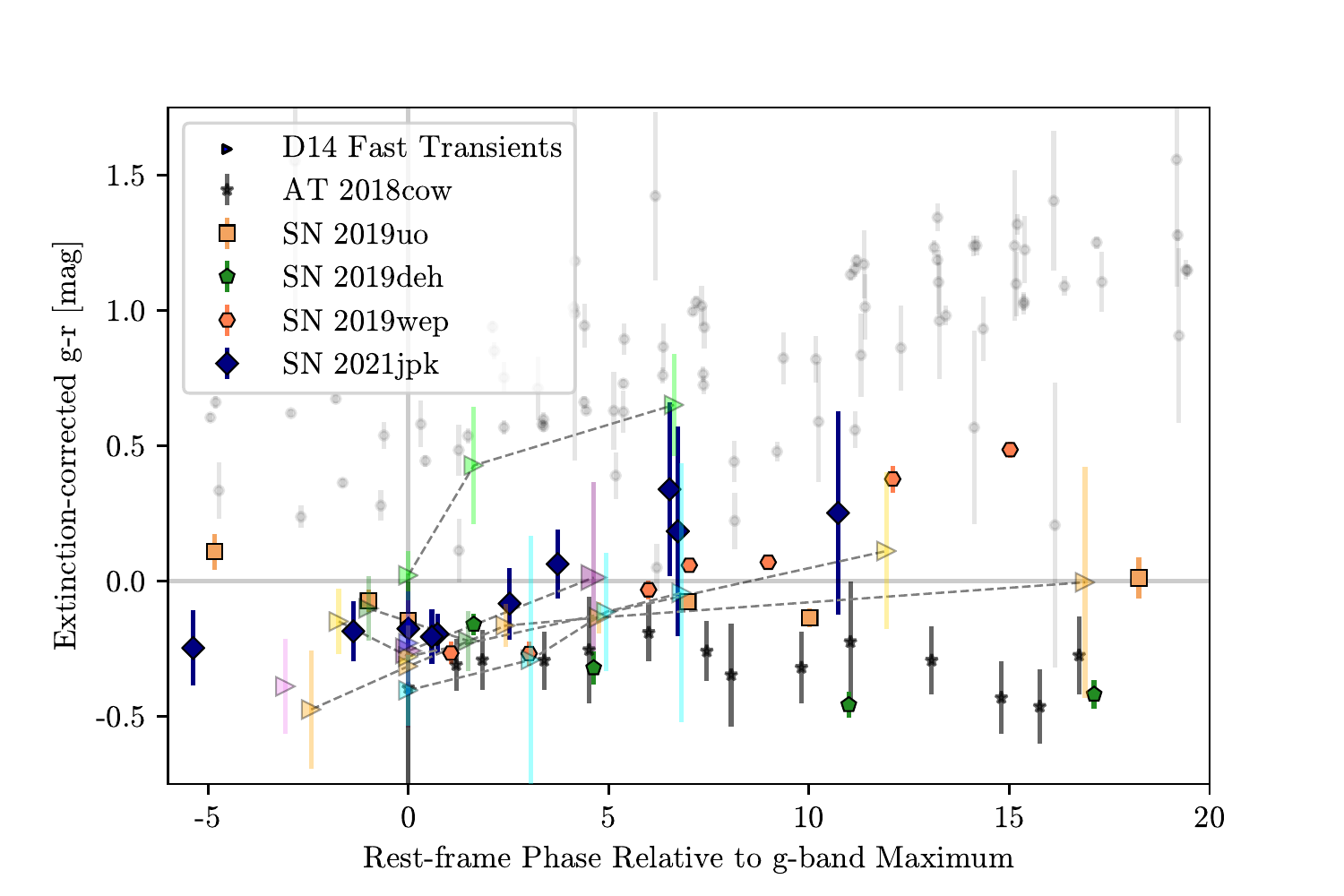}
    \caption{\textit{g-r} colors of the fast transients from D14 (triangles) compared with colors of fast-evolving SNe Ibn (colored shapes), AT\,2018cow (black stars), and a sample of SNe Ibc \citep[][gray points]{Taddia2015,Sako2018}. Colors have been corrected for MW extinction. PS1 fast transient colors are connected with dashed lines. The fast transients and SNe Ibn have colors that are mostly bluer than those of the comparison objects, particularly at later times.}
    \label{fig:colors}
\end{figure*}

\subsection{Photometric Properties}\label{subsec:phot}

The \textit{r}-band absolute magnitude light curves for the objects described in Sections \ref{subsec:lcodata} and \ref{subsec:cow} are shown in Figure \ref{fig:absmags}. Also included as the green-shaded region is the SN Ibn light-curve template presented in H17. Our objects show a wider range of peak luminosities and evolution timescales than the H17 template. For instance, SN\,2019uo has a lower peak absolute magnitudes and faster rise time than the template. However, \citet{Hosseinzadeh2017} state that because nondetections were not included in the fitting process, the template is biased to a brighter and shallower evolution at early times by SNe Ibn with longer rise times. AT\,2018cow is similar to the template in terms of peak \textit{r}-band absolute magnitude (-19.82 $\pm$ 0.06 and -19.46 $\pm$ 0.32 mag, respectively) and decline rate ($\approx$ 0.2 mag day$^{-1}$ and 0.1 mag day$^{-1}$, respectively), and is a closer match to SN\,2019deh, but it displays a brighter peak absolute magnitude by almost two magnitudes and a faster decline than the other SNe Ibn. These objects show that some SNe Ibn have rise times and decline rates that are more similar to those of luminous fast transients such as AT\,2018cow than other SNe Ibn, indicating that objects like AT\,2018cow may lie at the extreme end of a distribution of SNe Ibn.

In Figure \ref{fig:colors} we show the \textit{g-r} color evolution of these objects as well as that of the fast transients from D14. The colors of all the transients are presented in the observer frame. Although many of the D14 objects are at high redshifts, where \textit{K}-corrections become important, Figure 9 of D14 shows that \textit{K}-corrections do not greatly change the maximum light \textit{g-r} colors. Therefore we expect that \textit{K}-corrections will not significantly affect our conclusions here. In order to compare these colors with those of other types of SNe, we also plot the \textit{g-r} colors of a sample of Type Ib and Ic SNe \citep[SNe Ibc;][]{Taddia2015}. 

Before maximum light both the SNe Ibn and fast transients have consistently blue colors. After maximum light, some of these objects evolve to slightly redder colors. However, two objects\textemdash SN\,2019deh and AT\,2018cow\textemdash stand out as having little redward evolution when compared with the other objects. These extreme blue colors are rare, even among our sample of fast-evolving objects, and are evidence for photospheric temperatures that remain high throughout the evolution of these objects \citep{Drout2014,Perley2019}. More broadly, compared to the sample of SNe Ibc, the SNe Ibn and fast transient colors are mostly bluer at all times. One fast transient\textemdash PS1-12bb\textemdash is significantly redder than the other sample objects. However, as discussed in Section \ref{sec:csmmodels} PS1-12bb also has a different luminosity evolution than other fast transients and therefore may be an unrelated object.

\subsection{Blackbody Radius Measurements}

\begin{figure}
    \centering
    \includegraphics[width=0.45\textwidth]{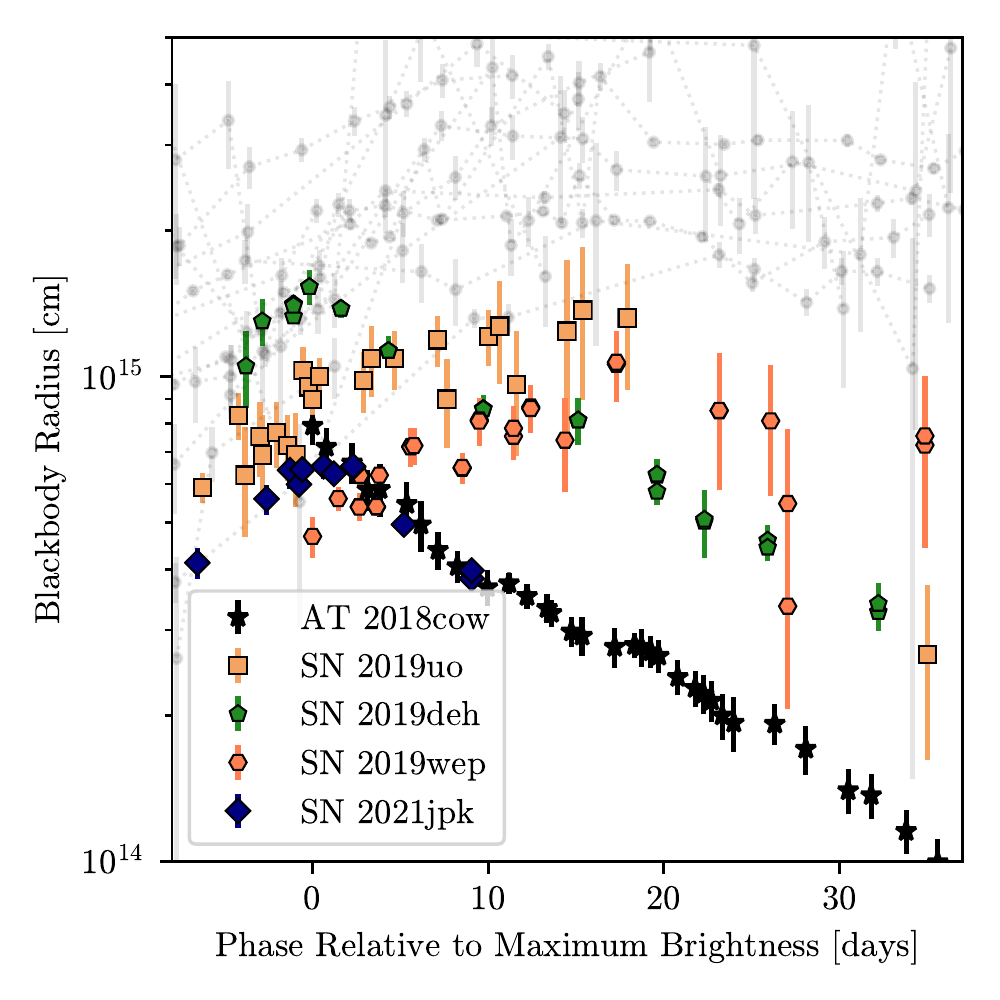}
    \caption{Blackbody radius measurements for our fast-evolving SNe Ibn as well as AT\,2018cow (from \citealt{Perley2019}) compared to estimates from a sample of SNe Ibc \citep[][gray points]{Taddia2015}. The SNe Ibn and AT\,2018cow have smaller blackbody radii at all epochs than the comparison objects, with a different evolution: our sample shows constant or decreasing blackbody radii after maximum light, whereas the others have constant or increasing radii.}
    \label{fig:bbradii}
\end{figure}

Given our multiband follow-up we are able to construct bolometric light curves for our sample of fast-declining SNe Ibn. For objects with no bolometric luminosity measurements published we fit a blackbody SED to our multiband photometry using the code \texttt{Superbol} \citep{Nicholl2018}. After correcting for MW extinction and shifting to the rest frame, we interpolate our observations to common epochs and fit for bolometric luminosities and blackbody radii and temperatures. We believe a blackbody approximation is valid as the spectra of the objects we consider are well modeled by blackbodies throughout their evolution. In order to ensure sufficient coverage in the UV, where the SEDs of these objects peak \citep{Drout2014}, we take care to measure luminosities only at epochs close to those with Swift observations. In the case of SN\,2019deh, only two UV observations were obtained, both around maximum light. In order to calculate the bolometric luminosity at later times we estimate magnitudes in the Swift UVOT filters by interpolating our LCO \textit{U}-band measurements onto a grid of \textit{uvw2-U}, \textit{uvm2-U}, and \textit{uvw1-U} colors from the archetypal Type Ibn SN\,2006jc \citep{Pastorello2007,Bianco2014, Brown2014}. Although an approximation, this method avoids the assumption of constant UV colors at later times which may lead to overestimated luminosities.

The blackbody radius evolution for our fast-evolving SNe Ibn and AT\,2018cow are shown in Figure \ref{fig:bbradii}. For comparison, we plot blackbody radii estimates from a sample of SNe Ibc \citep{Taddia2015, Sako2018}. \citet{Ho2021} notice that AT\,2018cow and some SNe Ibn are distinct from other transients in that their blackbody radii decrease over time. We find a similar trend for most of our objects. All have blackbody radii of $\approx$ $10^{14}$--$10^{15}$ cm that tend to plateau or decrease after maximum light. The faster-evolving objects tend to have decreasing radii, whereas the radii of the slower-evolving SNe are more constant. These properties are distinct when compared to the SNe Ibc, which display larger radii that remain constant or increase after maximum light. AT\,2018cow is somewhat unique as it has a smaller peak blackbody radius than the SNe Ibn, with the exception of SN\,2021jpk. This may be evidence of a more confined CSM, as discussed in Section \ref{sec:discussion}.

\subsection{Spectroscopic Evolution}\label{subsec:spec}

\begin{figure*}
    \centering
    \subfigure[]{\includegraphics[width=0.8\textwidth]{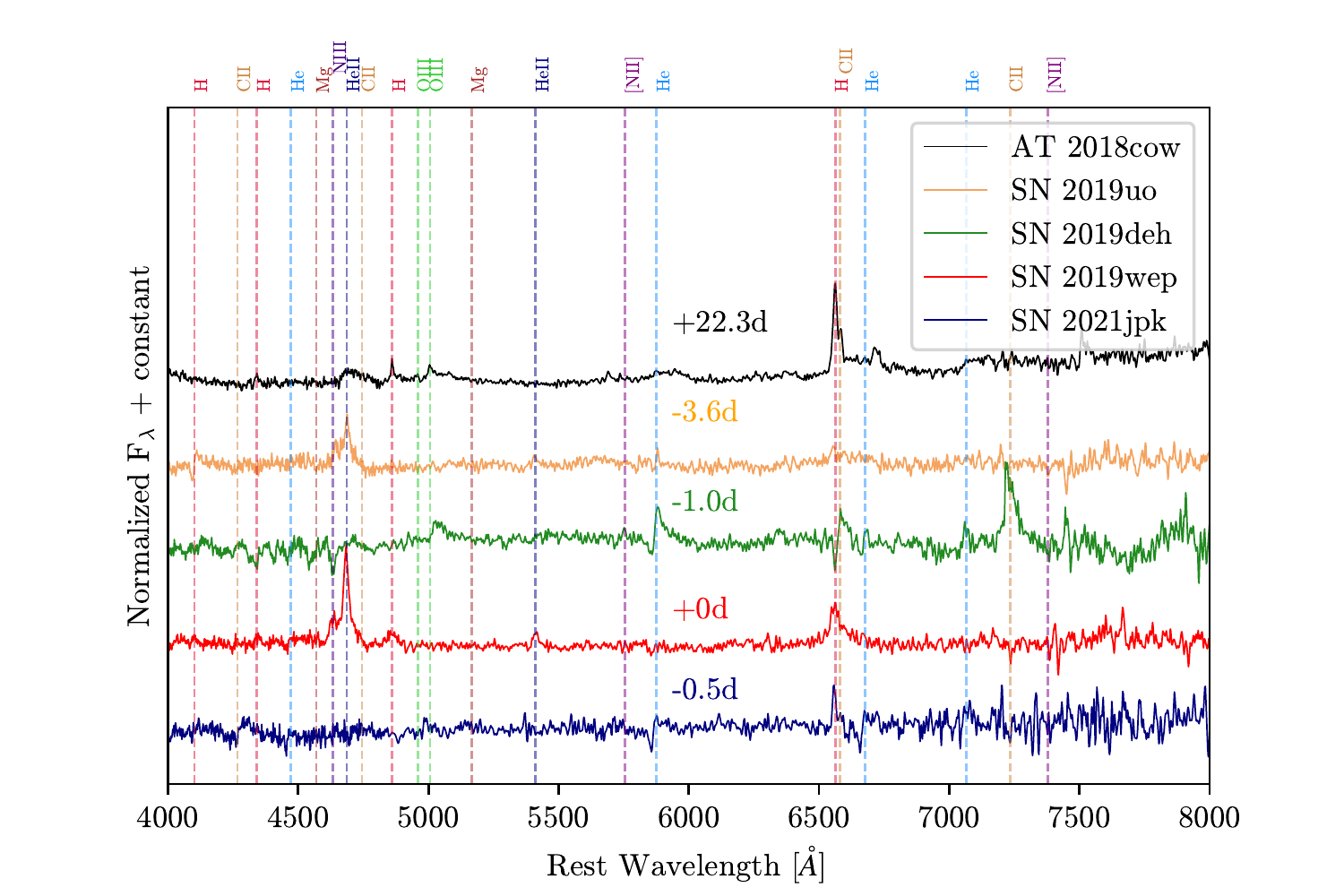}}\label{fig:speclate}
    \subfigure[]{\includegraphics[width=0.8\textwidth]{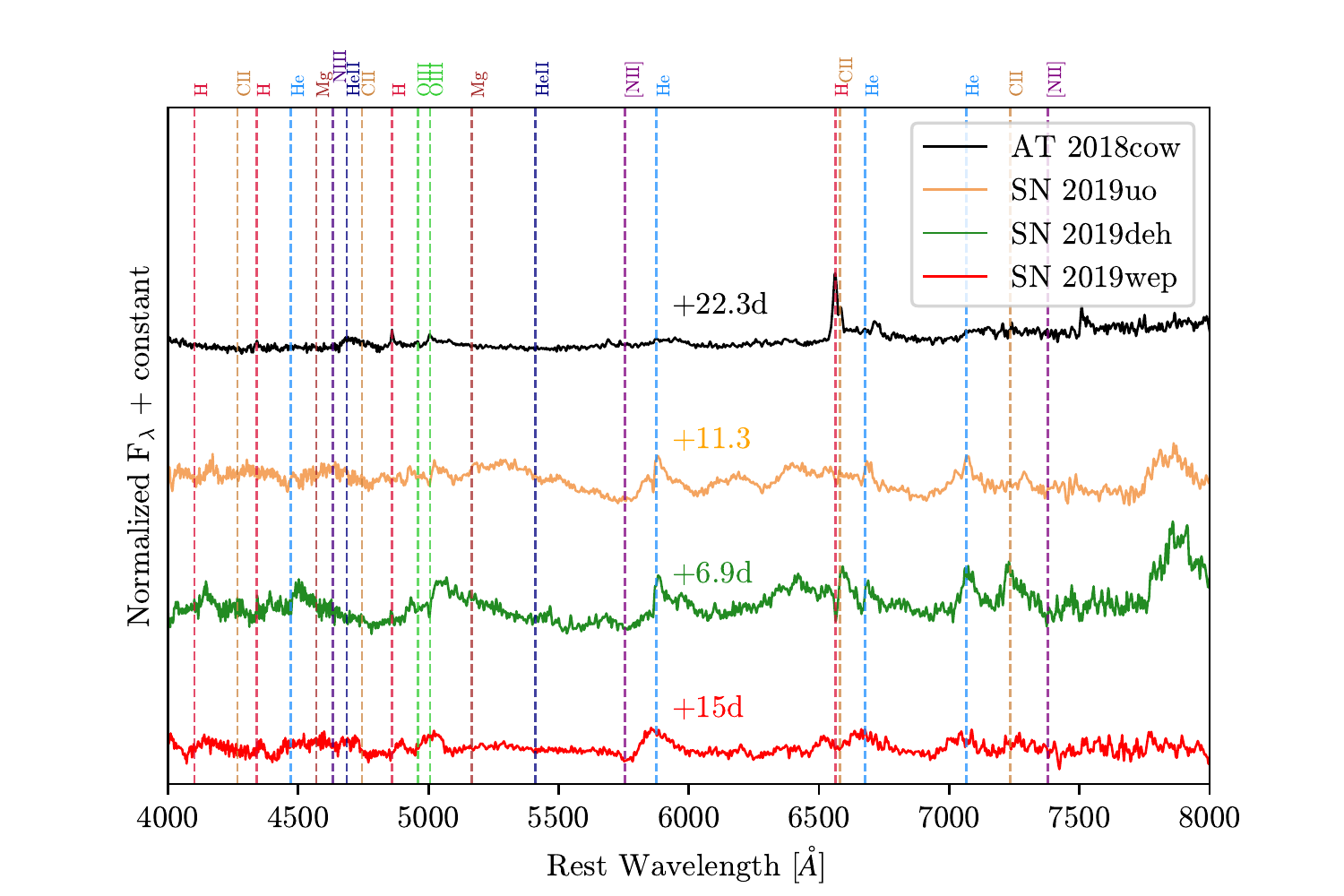}}\label{fig:specearly}
    \caption{(a) A comparison of the early-time spectra of the fast-evolving SNe Ibn to that of AT\,2018cow at a later phase. All spectra have been continuum subtracted. Phases with respect to \textit{g}-band peak brightness are labeled for each spectrum and spectral features are marked with dotted lines. The SNe Ibn spectra are similar to that of AT\,2018cow, hinting that SNe Ibn-like spectral features in AT\,2018cow may be hidden at earlier times. (b) Same as above, but here the spectra of the SNe Ibn are one to two weeks past maximum. At this stage the features of the SNe Ibn are more developed than those of AT\,2018cow. SN\,2021jpk is not included as spectra were only obtained at maximum light.}
    \label{fig:spec}
\end{figure*}

A defining characteristic of fast transients is their featureless blue continua, which makes spectroscopic classification difficult \citep{Drout2014}. AT\,2018cow is one such object, with featureless spectra closely approximating a blackbody for the first $\approx20$ days of its evolution \citep{Prentice2018}. Blue, featureless continua are often found in young core-collapse SNe, in which the expanding ejecta is still hot and optically thick, preventing the formation of P-Cygni features \citep[e.g.,][]{Hosseinzadeh2017}. A constant featureless continuum is evidence for a sustained powering mechanism such as shock cooling or CSM interaction.

Similarly, at early times SNe Ibn have spectra that show hot blue continua superimposed with narrow lines of He and other elements. These narrow lines originate from highly ionized species in a nearby CSM shell or wind and disappear once this material recombines or is swept up by the SN explosion, providing direct observable evidence for CSM interaction in SNe Ibn. It has been noted that AT\,2018cow shows similar signs of interaction with a He-rich CSM. \citet{Fox2019} find that the spectra of AT\,2018cow are qualitatively similar to those of SNe Ibn convolved with a hot ($\sim 10^4$K) blackbody. Additionally, similar spectral features, including narrow- and intermediate-width He emission lines from pre- and post-shocked CSM, are seen in both the late-time spectra of AT\,2018cow and in spectra of SNe Ibn \citep{Fox2019,Xiang2021}.

Figure \ref{fig:spec} compares an LCO spectrum of AT\,2018cow to LCO spectra of fast-evolving SNe Ibn at both early (top) and late (bottom) times. The spectra have been normalized as follows: from each spectrum we subtract the flux from its best-fit blackbody and divide by the median of the remaining flux. We find that a blackbody fits the continua well for all phases we consider here. When comparing a later spectrum of AT\,2018cow with spectra of SNe Ibn around maximum light, the objects show qualitative similarities. All the objects show hot blue continua before normalization with emission lines of He, including a He II emission line in the spectrum of AT\,2018cow that is broader than the same line in the spectra of SN\,2019uo and SN\,2019wep. Additionally, at early times SN\,2019uo and SN\,2019wep show flash features of C III and N III \citep[][Gangopadhyay 2021, in preparation]{Gangopadhyay2020}. However, at about two weeks after maximum the spectra of SNe Ibn show more developed emission and P-Cygni features of He I, C II, and O III than AT 2018cow, which still resembles a hot blackbody with few narrow emission lines. This consistently high photospheric temperature may be evidence of a long-lasting powering source for AT\,2018cow, such as sustained CSM interaction, which can mask the underlying spectrum \citep{Fox2019}.

Despite their different evolution at later times, the similar spectral features between AT\,2018cow and the SNe Ibn, including a strong blue continuum and emission lines of He I and He II, hint at a common progenitor system. In particular, narrow emission lines with WR-like features are evidence for CSM interaction \citep{Taddia2013,GalYam2014}, pointing to a common circumstellar environment. In the case of SNe Ibn, it is unclear if the CSM has a shell-like density profile due to a preexplosion outburst in the months or years before explosion \citep[e.g.,][]{Smith2008} or a wind-like profile from a WR stellar wind (see e.g., \citealt{Crowther2007} for a review). AT\,2018cow shows more peculiarities than SNe Ibn, such as He I emission features in the late-time spectra that are redshifted by several thousand km s$^{-1}$ \citep{Benetti2018}. This redshift may be explained by asymmetries in a CSM shell \citep{Margutti2019}, which may be common in interaction-powered SNe \citep{Soumagnac2020}. Additionally, the emission features are broader and do not appear until much later than in SNe Ibn. This different evolution at later times may be the case if the CSM is closer to the progenitor star and is quickly overrun by the optically thick ejecta at early times \citep{Fox2019}. Once the ejecta has expanded and cooled, the optical depth will drop and broadened emission lines from continued interaction with post-shocked CSM can be observed \citep{Fox2019}. This may imply that the CSM is much more confined in radius in the case of AT\,2018cow than in SNe Ibn. A discussion of the CSM properties of these objects is given in Section \ref{subsec:modelfits}.

\section{Circumstellar-interaction Models} \label{sec:csmmodels}

\subsection{Model Description}\label{subsec:modeldesc}

Modeling the energy source powering the light curves of SNe Ibn and other fast transients is necessary to understanding their progenitor systems. Due to their similar colors, photometric evolution, and spectral features, it is plausible that photometrically classified fast transients have a similar powering mechanism and progenitor environment to SNe Ibn. In SNe Ibn, the combination of narrow emission lines seen in spectra at early times and the fast rise to peak luminosity point to CSM interaction as a primary power source. Modeling the light curves of SNe Ibn has shown that either a combination of CSM interaction and $^{56}$Ni decay \citep{Clark2020,Wang2020,Wang2021} or CSM interaction alone \citep{Karamehmetoglu2019} can sufficiently reproduce their luminosity evolution. 

Due to their rarity \citep[$\approx$ 0.1$\%$ of the core-collapse SNe rate for AT\,2018cow-like transients;][]{Ho2021} and rapid evolution, the mechanisms powering the light curves of fast transients have not been as well studied. However, radioactive decay cannot be the sole powering mechanism, as the amount of radioactive Ni needed to reach high peak luminosities in only a few days often exceeds the total ejecta mass by an order of magnitude \citep{Drout2014,Arcavi2016,Pursiainen2018,Rest2018}. Multiple physical interpretations of AT\,2018cow have been suggested, including powering due to a central engine \citep{Prentice2018,Margutti2019}, shock breakout from an optically thick shell of CSM \citep{RiveraSandoval2018}, and the tidal disruption event of a white dwarf \citep{Perley2019}. More recently, \citet{Xiang2021} modeled the bolometric light curve of AT\,2018cow with CSM interaction and $^{56}$Ni decay. This choice is physically motivated by the similar luminosity evolution between AT\,2018cow and SNe Ibn as well as the narrow emission lines of He and C seen in its spectra, as described in Section \ref{sec:data}. They found that the light curve of AT\,2018cow can be reasonably explained by an energetic explosion with a small amount of ejected mass within an optically thick CSM shell or wind of small inner radius. 

To explore whether the same powering source can sufficiently reproduce the light-curve evolution of both SNe Ibn and other fast transients, we construct a grid of CSM interaction plus $^{56}$Ni decay models. We begin with the code presented in \citet{Jiang2020}, which finds self-similar solutions to the interaction between expanding SN ejecta and a stationary CSM as first presented by \citet{Chevalier1982} and \citet{Chevalier1994}. This model assumes a two-zone SN ejecta: an inner region with a shallow density profile, $\rho \propto r^{-\delta}$, and an outer region with a much steeper profile, $\rho \propto r^{-n}$. The CSM density is parametrized as $\rho \propto r^{-s}$, with $s=0$ being a shell-like CSM and $r=2$ being a wind-like CSM. To this solution we also add the analytic formalism for $^{56}$Ni decay with diffusion as presented in \citet{Chatzopoulos2012}.

\begin{figure*}
    \centering
    \subfigure[]{\includegraphics[width=0.8\textwidth]{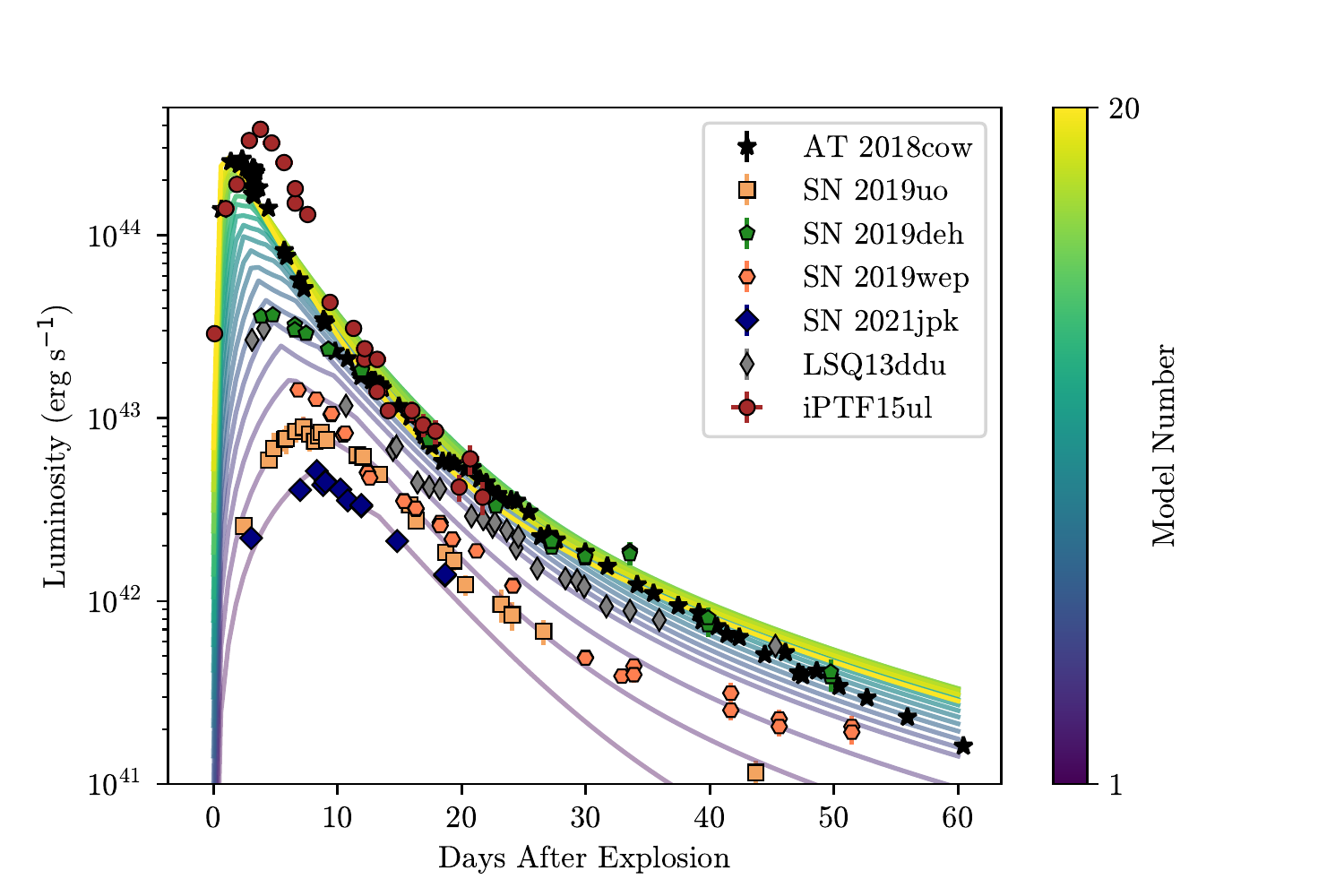}}\label{fig:modelswithibns}
    \subfigure[]{\includegraphics[width=0.8\textwidth]{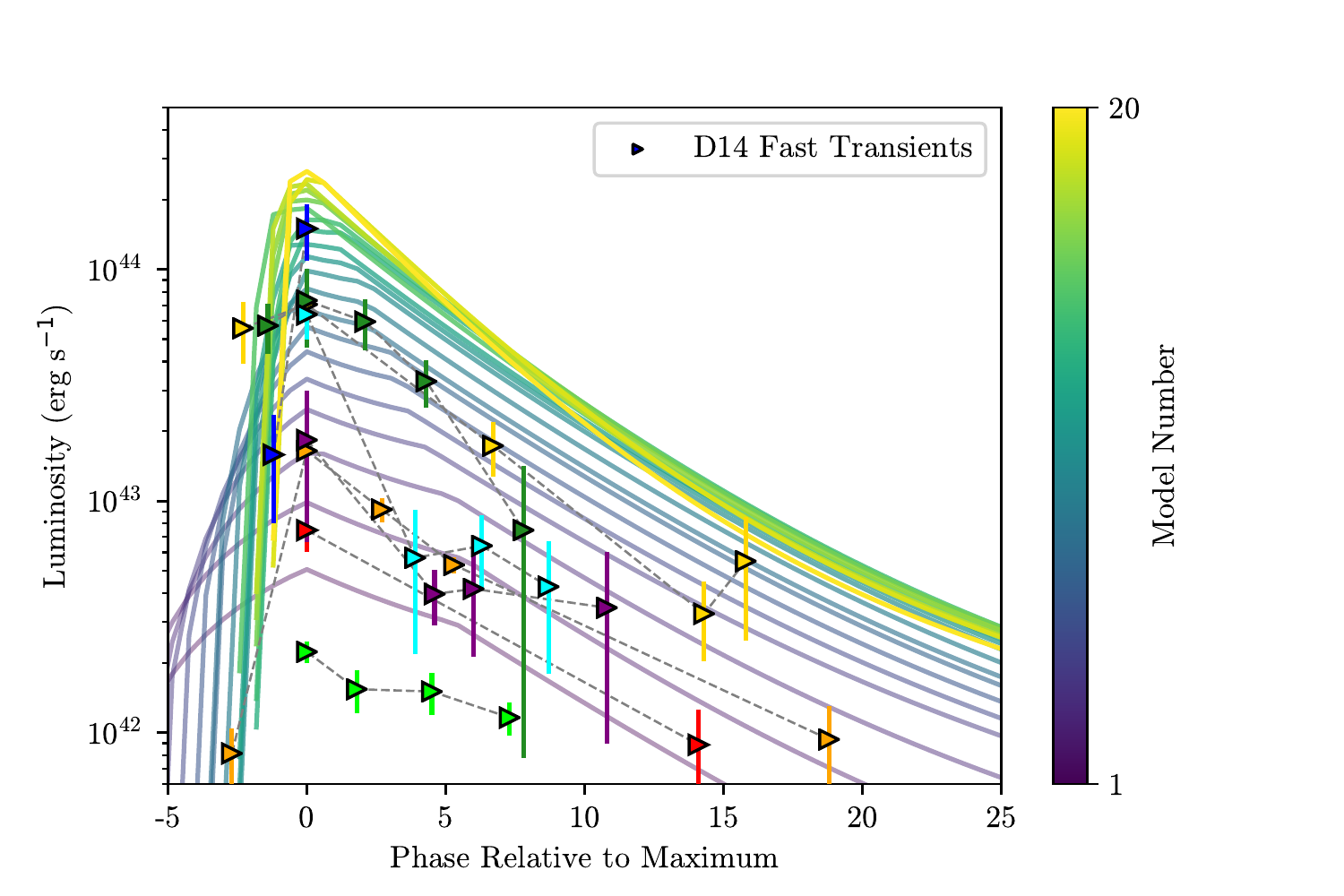}}\label{fig:modelswithfbots}
    \caption{(a) Model light curves powered by CSM interaction and $^{56}$Ni decay (solid curves) compared with bolometric luminosities of fast-evolving SNe Ibn (colored points) and the fast transient AT\,2018cow (black stars). Model parameters are given in Table \ref{tab:modelparams}. The model light curves span the range of luminosities between the faint, fast-evolving SNe Ibn and AT\,2018cow. (b) Same as the top figure, but comparing the model light curves to the fast transients from D14. Again the models replicate the luminosity evolution of many of the objects. Note here that phase is plotted with respect to the time of \textit{g}-band maximum.}
    \label{fig:model_lcs}
\end{figure*}

Because this CSM interaction plus $^{56}$Ni decay model has the potential for degeneracy between its many input parameters, we first focus on qualitatively reproducing the observed evolution of the SNe Ibn and fast transient light curves in order to gain a better understanding of the progenitors of these objects. To do so, the model is fit to the bolometric light curves, calculated from our UV and optical photometry, of our faintest objects (SN\,2021jpk and SN\,2019uo) and our brightest (AT 2018cow) using the Markov Chain Monte Carlo routine \texttt{emcee} \citep{ForemanMackey2013}. This is done in order to obtain initial model parameters that reproduce the light curves of both the faint, slower-evolving and bright, fast-evolving objects. These initial parameters are compared with best-fit values published for the objects in our sample \citep{Gangopadhyay2020,Xiang2021} as well as other SNe Ibn and fast transients \citep[e.g.,][]{Karamehmetoglu2019,Clark2020}. We find reasonable qualitative agreement between our values and those published in the literature. In order to reproduce the light curves of all the objects in our sample, we smoothly vary the model parameters between the initial values of the faint, slow-evolving objects and the bright, fast-evolving ones. The following parameters are varied: 
\begin{enumerate}
    \item $v_{\text{ej}}$, the ejecta velocity;
    \item $M_{\text{ej}}$, the ejecta mass;
    \item $M_{\text{CSM}}$, the CSM mass;
    \item $R_0$, the inner radius of the CSM;
    \item $\rho_0$, the density of the CSM at the inner radius;
    \item $\epsilon$, the radiation efficiency;
    \item $\kappa_{\gamma}$, the gamma ray opacity; and
    \item $M_{\text{Ni}}$, the mass of $^{56}$Ni produced.
\end{enumerate}
For all the models we set $n=10$, $\delta=1$, $s=0$, and the optical opacity $\kappa_{\rm opt} = 0.1$ cm$^{2}$ g$^{-1}$. We caution that allowing these parameters to vary as well may also provide good fits to the data. This model grid is not meant to produce best fits to the data (for our efforts to find best fits, see Section \ref{subsec:minimmodels}). Instead, it is simply meant to illustrate similarities between the progenitor environments of SNe Ibn and fast transients and show that a continuous range of initial conditions can reproduce the behavior of both classes of objects. The full list of parameter values for the 20 models presented is given in Table \ref{tab:modelparams}.

\subsection{Comparison to Observations}\label{subsec:modelfits}

Our model light curves are shown in Figure \ref{fig:model_lcs}. Different colors correspond to different models with parameters shown in Table \ref{tab:modelparams}. The top panel compares the models to fast-evolving SNe Ibn as well as AT\,2018cow. We have supplemented our sample of SNe Ibn with published data of LSQ13ddu \citep{Clark2020} and iPTF15ul \citep{Hosseinzadeh2017}. The models span over 1.5 orders of magnitude between the faintest object (SN\,2021jpk) and the most luminous (iPTF15ul). Additionally, they reproduce the observed rise and decline times of these objects, including the rapid evolution of AT\,2018cow, as well as the luminosity evolution of most of the objects out to $\sim$60 days after explosion. 

The bottom panel of Figure \ref{fig:model_lcs} compares the same models with the fast transients in D14. The range of luminosities is again reproduced by the models. Due to the small number of observations at similar epochs, the light curves of the objects are more sparsely populated, which makes comparing their late-time evolution to the models difficult. At least some of the objects have similar rise times and decline rates as the models, whereas others show different evolution. However, several factors make comparing this sample to the CSM models difficult. First, the PS1 objects have poorly constrained phases and late-time evolution due to their sparse light-curve sampling. Additionally, several of the objects show a tentative increase in the bolometric luminosity roughly ten days after maximum light. \citet{Ho2021} suggest these objects may be Type IIb SNe. In these cases, the observed rapid decline may be caused by shock-cooling emission, while a low $^{56}$Ni mass may produce a weak secondary peak that went unobserved. Besides these cases, however, the broad agreement between the luminosities and decline rates of our models and the PS1 fast transients suggests that some of these objects may be powered by CSM interaction.

\begin{figure*}[t!]
    \centering
    \includegraphics[width=0.8\textwidth]{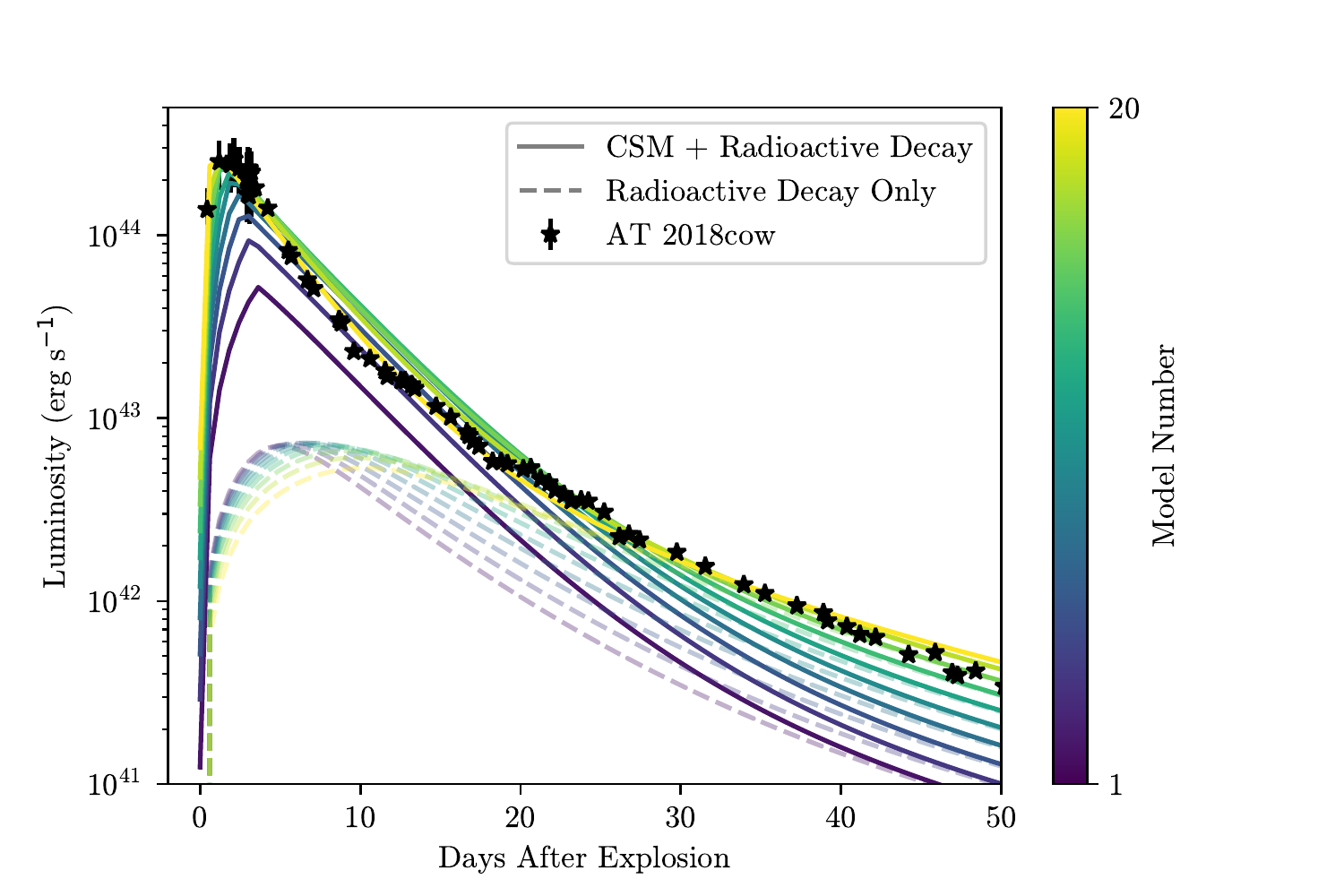}
    \caption{Model light curves with fixed ejecta parameter values but varying CSM parameter values. Plotted for comparison is the bolometric light curve of AT\,2018cow. The radioactive decay component of each light curve is plotted with a dashed line. Model numbers correspond to the CSM parameter values from Table \ref{tab:modelparams}.}
    \label{fig:modelvaryingcsm}
\end{figure*}

To test the effect the CSM parameters have on the models, we construct a separate grid with the same $M_{\text{ej}}$, $v_{\text{ej}}$, $\kappa_\gamma$, and $M_{\text{Ni}}$ values as our model that best matches AT\,2018cow, but with CSM parameters that are varied over the range of values in Table \ref{tab:modelparams}. The results are shown in Figure \ref{fig:modelvaryingcsm}. We find that models with smaller and denser CSM shells power light curves that evolve faster and reach higher peak luminosities. The radiation diffusion timescale is also significantly impacted by the choice of CSM parameters. For AT\,2018cow, this leads to a transition from CSM interaction to radioactive decay as the primary powering mechanism at $\approx$ 20 days after peak, as noted previously \citep{Xiang2021}.

\begin{figure*}
    \centering
    \subfigure[]{\includegraphics[width=0.48\textwidth]{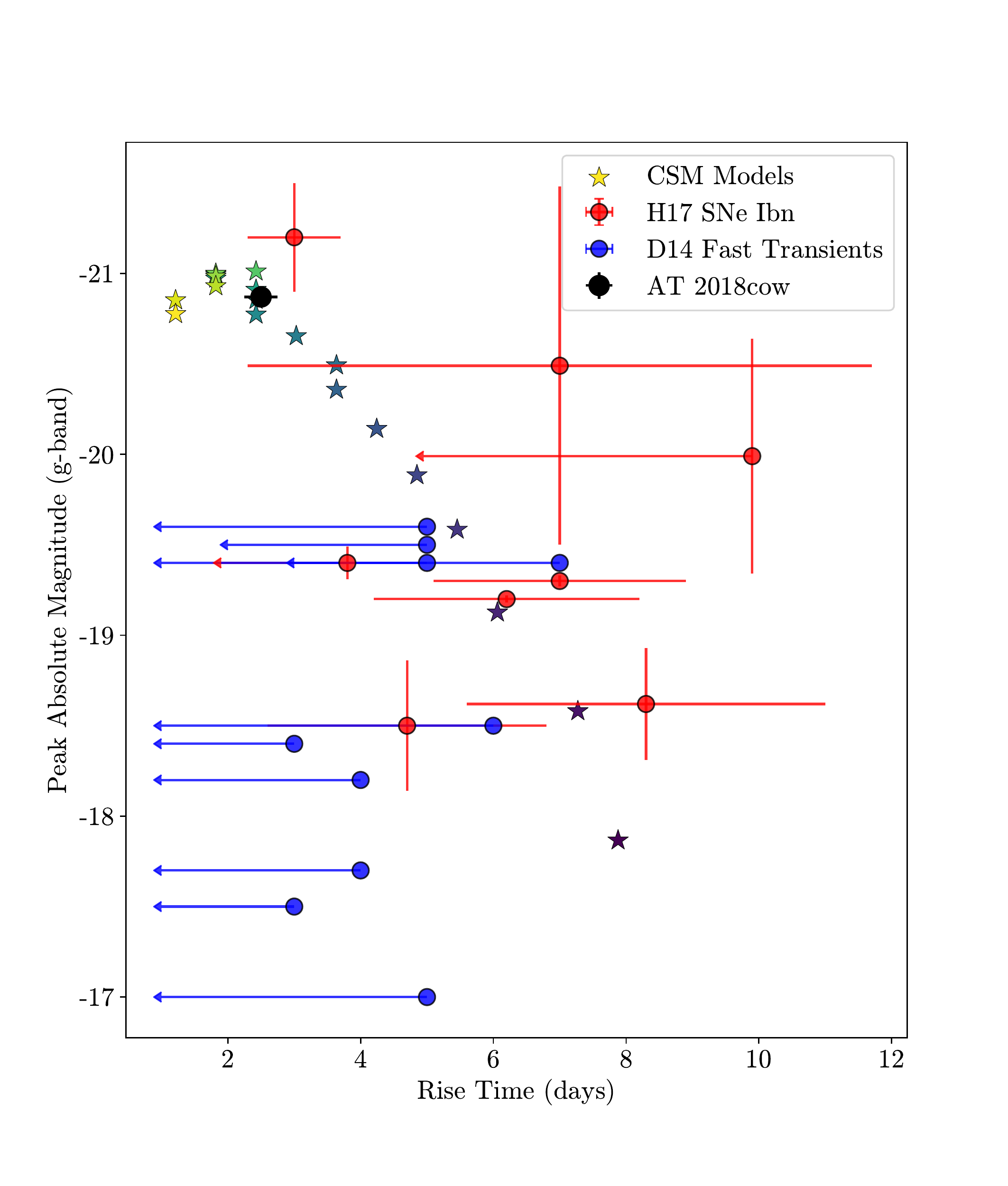}}\label{fig:trise}
    \subfigure[]{\includegraphics[width=0.48\textwidth]{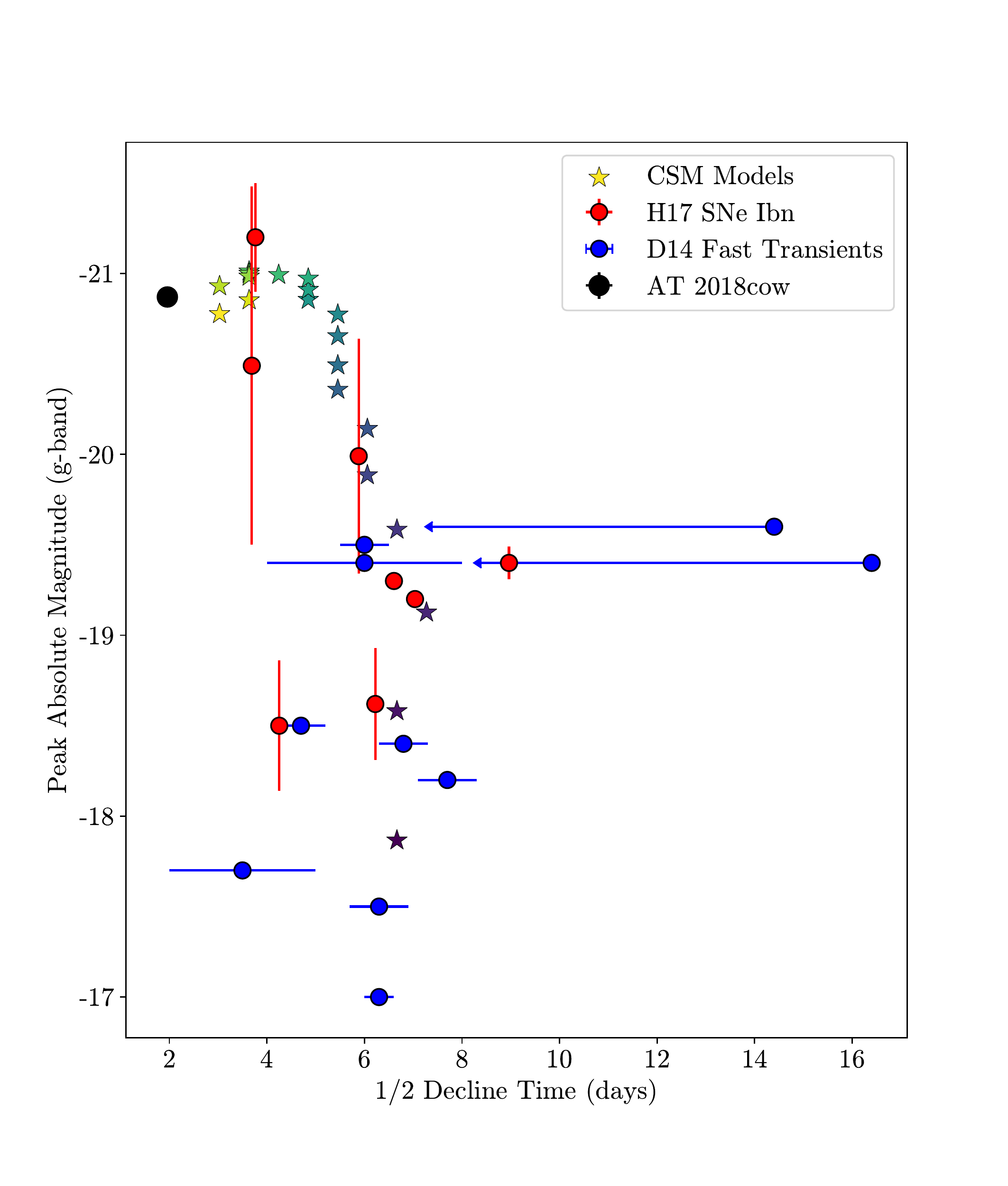}}\label{fig:tdec}
    \caption{(a) Rest-frame \textit{g}-band peak absolute magnitude versus rise time for our model light curves (colored stars) compared to SNe Ibn from H17 (red points), fast transients from D14 (blue points), and AT\,2018cow (black point). The color map is the same as in Figure \ref{fig:model_lcs}. The models span the parameter space between the fast-evolving SNe Ibn and the luminous fast transients. The fast transients not matched by these models are discussed in Section \ref{subsec:csmsufficient}. (b) Peak absolute magnitude versus time to decline by half the peak luminosity for the same objects. Again, the models span the parameter space between SNe Ibn and fast transients.}
    \label{fig:tparam}
\end{figure*}

In order to compare these models with the larger sample of objects in a different parameter space, in Figure \ref{fig:tparam} we plot the rest-frame peak \textit{g}-band absolute magnitudes versus rise times and decline times of the models, several of the SNe Ibn in H17, the fast transients in D14, and AT\,2018cow. To estimate peak absolute magnitudes of the model light curves, we find the flux within the \textit{g}-band assuming a blackbody SED given by the photospheric radius and temperature of each model. 

In this phase space the model light curves exist on the boundary between the D14 fast transient and SNe Ibn populations. Additionally, the more luminous models evolve the fastest, which is key to replicating the behavior of the brightest, fastest-evolving transients such as AT\,2018cow. Although there are still objects in this parameter space that are not matched by the models, we have reproduced the range of light-curve behaviors of the more luminous fast-evolving objects, including SNe Ibn, many of the fast transients in D14, and more extreme objects such as AT\,2018cow. This shows that SNe Ibn and some other fast transients may share a common powering source, rather than having distinct physical mechanisms.

\subsection{Comparison with Other Model Results}\label{subsec:minimmodels}

In order to test the model dependency of these results, we fit the bolometric light curves of all the objects shown in Figure \ref{fig:model_lcs} utilizing the {\tt Minim} code and applying the {\tt hybrid} model presented in \citet{Chatzopoulos2013}. This model assumes the interaction between an optically thick CSM and an expanding SN ejecta, using the self-similar solution of \citet{Chevalier1982} for the calculation of the expansion of the forward and reverse shocks, as the main powering mechanism. The shock heating efficiency, $\epsilon$, is assumed to be 100$\%$. While this may not be fully true in reality, this assumption provides a useful lower limit for the strength of the CSM interaction without introducing an additional (poorly constrained) parameter for the shock heating efficiency. In addition, the usual radioactive Ni-Co-Fe decay is used as the heating source of the SN ejecta. The CSM is modeled as a simple, constant-density shell with an inner radius of $R_{\rm ej}$ and an outer radius specified by its mass ($M_{\rm CSM}$) and density ($\rho_{\rm CSM}$). As earlier, the density structure of the SN ejecta is assumed as an outer power law, this time having $n=12$ (a built-in value in the {\tt hybrid} model) and an inner, flat region within $r_0 = 0.1 R_{\rm ej}$. Note that instead of the CSM density, the formal mass-loss rate
$\dot{{\rm M}} = 4 \pi {\rm R}^2_{\rm ej} \rho_{\rm CSM} v_{\rm w}$ with $v_{\rm w} = 10$~km~s$^{-1}$ is used in the {\tt hybrid} model as a fitting parameter, even though it has no direct physical meaning in the context of a constant-density CSM cloud. After the forward and reverse shock passes through the CSM shell and the ejecta, the shock-heated material radiates out its thermal energy via radiative diffusion. The {\tt hybrid} model applies the usual constant-opacity approximation. We set the optical opacity as $\kappa_{\rm opt} = 0.1$ cm$^2$~g$^{-1}$ and the gamma-ray opacity as $\kappa_\gamma = 0.03$ cm$^2$~g$^{-1}$.

The best-fit models are selected based on $\chi^2$ minimization by applying the Price algorithm, which samples the parameter space with a controlled random-search method \citep[see][for more details]{Chatzopoulos2013}. Parameters of the best-fit models are collected in Table~\ref{tab:minimfit}. The parameters are in reasonable (order-of-magnitude) agreement with those shown in Table \ref{tab:modelparams}, despite the different assumptions made between the two models (including different ejecta power-law indices and efficiency values) and the fact that the CSM models presented earlier were produced to qualitatively match the range of observed light curves properties without performing rigorous best-fit routines. This agreement supports the insight our model grid gives into the progenitor systems of these fast-evolving objects.

Our calculations using the {\tt Minim} code reveal that due to the rapid light-curve evolution of these transients, both the forward and reverse shocks sweep up the CSM and the SN ejecta by approximately the time of maximum light. After maximum, the decline of the light curve can be explained by the cooling of a shock-heated ejecta and CSM. This behavior is different from what is observed in other interacting (Type IIn) SNe, where the shocks live much longer and the CSM cloud stays optically thick on a longer timescale.

Our best-fit {\tt Minim} light curves are shown in Figure \ref{fig:minimfits}. We find that the {\tt Minim} models provide almost perfect fitting to the data of the SNe Ibn, assuming both CSM interaction and Ni-Co decay as coexisting heating sources; without the radioactive energy input, the CSM-only light curves are not compatible with the observations. On the other hand, AT\,2018cow is peculiar because its long-lasting quick decline rate is not well described by the predicted Ni-Co decay rate at late phases. For this object the CSM-only model can fit the peak of the light curve, but then the model light curve declines too fast, which suggests the presence of an additional heating source. Because the {\tt hybrid} model in {\tt Minim} assumes full trapping of the $\gamma$ rays from Ni-Co decay, it is possible that $\gamma$-ray leaking (possibly caused by noncentral Ni distribution or a nonspherical ejecta geometry) may explain the unusual decline of AT\,2018cow. We further discuss the possibility of an asymmetric ejecta for AT\,2018cow in Section 
\ref{subsec:csmsufficient}.

\section{Discussion}\label{sec:discussion}

\subsection{Is CSM Interaction Sufficient to Model Fast Transients?}\label{subsec:csmsufficient}

\begin{figure*}
    \centering
    \includegraphics[width=0.75\textwidth]{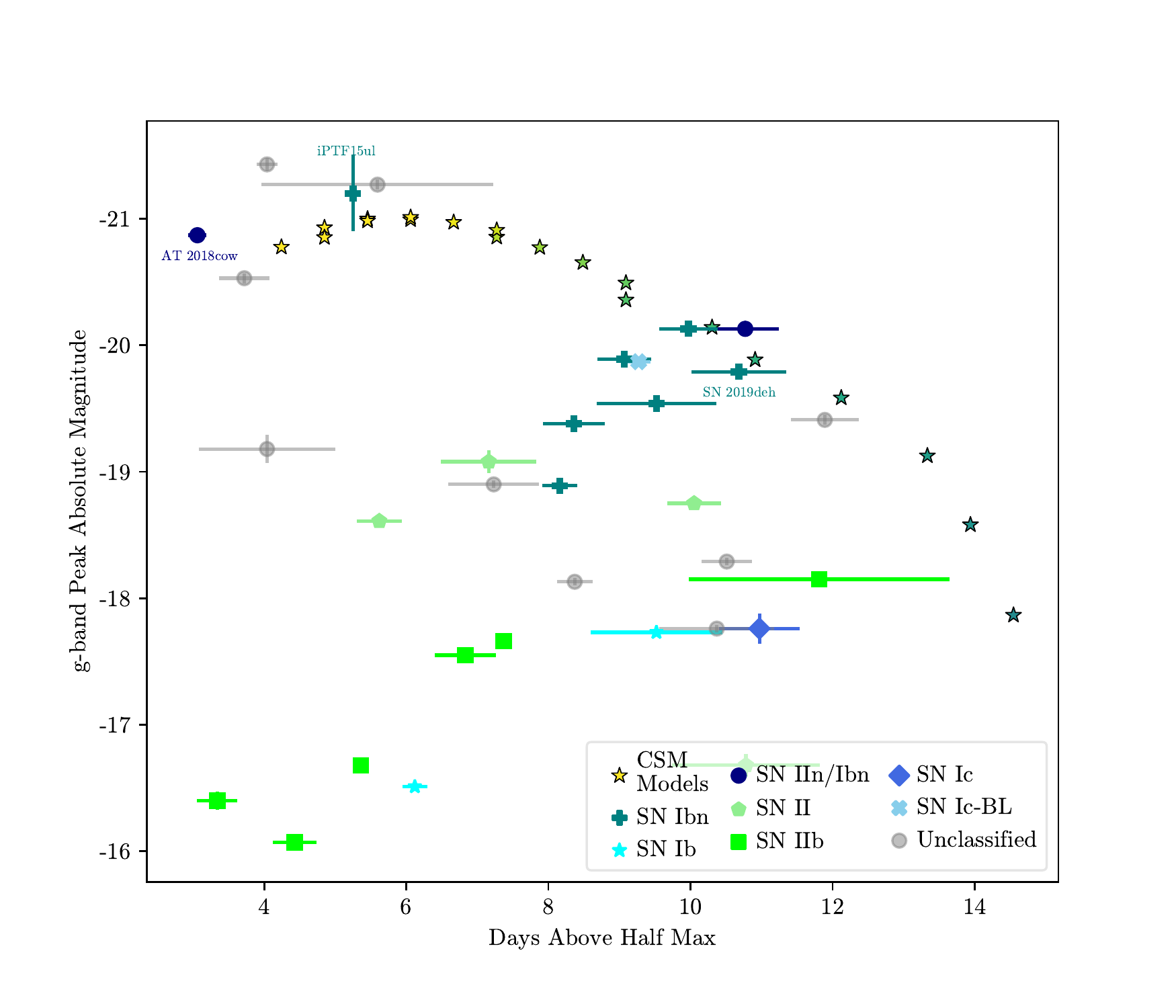}
    \caption{Rest-frame \textit{g}-band absolute magnitude versus time above half-maximum of our CSM models (colored stars) compared to that of several of the objects presented in \citet[][other symbols]{Ho2021}. Spectroscopically classified transients are plotted as colored symbols and three objects from \citet{Ho2021} that we discuss in this work are labeled.}
    \label{fig:ztf}
\end{figure*}

Based on the similarities discussed in Section \ref{sec:data}, we are motivated to consider a common powering source and progenitor system between SNe Ibn and some fast transients. These observational similarities include the following:

\begin{enumerate}
    \item a similar color evolution, with colors that are consistently bluer than other SNe Ibc;
    \item similar blackbody radius evolution, with both classes of objects exhibiting receding photospheres after peak brightness; and
    \item similar spectral features, such as a hot blue continuum superimposed with narrow He lines as well as occasional flash features of He, C, O, and other highly ionized elements.
\end{enumerate}

These common characteristics can all be explained by interaction between CSM and the SN ejecta. Modeling this interaction as the primary powering source of these objects at early times, along with a $^{56}$Ni decay component, we are able to reproduce the range of rise times, peak luminosities, and decline rates in our sample of SNe Ibn, some fast transients from D14, and AT\,2018cow.

However, these CSM interaction models do not match the observed light-curve properties of all the fast transients reported in D14. In particular, the model parameters we consider here are unable to reproduce the fainter fast-rising objects. \citet{Ho2021} studied a large sample of spectroscopically classified fast transients in ZTF Phase I and found that objects in this  region of parameter space (i.e. $t_{rise} \lesssim 5$ days and peak $M_g \gtrsim -18$) were mainly SNe IIb. It is more likely, therefore, that the faint and fast-rising transients in D14 are observed shock-cooling light curves from SNe IIb, and therefore are physically distinct from those powered by CSM interaction.

It is also possible that other powering mechanisms, such as a central engine, are needed to reproduce some observed features of fast transients, such as the high ejecta velocities and X-ray luminosities seen in AT\,2018cow and other similar transients \citep{Coppejans2020,Ho2020,Perley2021}. \citet{Ho2021} argue that the high radio and X-ray luminosity of AT\,2018cow and several other spectroscopically unclassified fast transients set them apart from other objects with rapid evolution, including luminous SNe Ibn. On the one hand, X-ray and radio emission can arise from CSM interaction \citep{Chevalier1982,Chevalier1994}. \citet{RiveraSandoval2018} initially use the variable X-ray luminosity of AT\,2018cow as evidence of CSM interaction powering the light curve. Their estimated CSM radii ($\approx$ 100 - 200 au) and masses ($\gtrsim0.08$ M$_\odot$), inferred from the X-ray emission are qualitatively similar to our model parameters. On the other hand, \citet{Margutti2019} argue against an external CSM shock as the primary power source for AT\,2018cow. Observations of the early X-ray luminosity of AT\,2018cow disfavor an external shock as the source of the X-ray emission and instead show the need for a central source of high-energy photons \citep{Margutti2019}. Similar X-ray luminous and/or radio-loud fast transients have recently been discovered at cosmological distances \citep{Coppejans2020,Ho2020,Perley2021}. If these transients have the same physical mechanism as AT\,2018cow, the central X-ray source must be physically distinct from the external interaction powering the luminous radio emission \citep{Ho2019}. 

However, \citet{Margutti2019} show that interior shocks originating from ejecta interacting with a dense equatorial CSM ring may be sufficient to power the X-ray luminosity observed in AT\,2018cow. A highly asymmetric CSM has been attributed to other astrophysical phenomena, including luminous red novae \citep{Metzger2017} and SNe such as iPTF14hls \citep{Andrews2018}, and may arise naturally from binary interaction \citep{Sana2012} or explosive mass loss \citep{Smith2014}. Asymmetries in the CSM may explain some of the other unusual features of this object. Most of the ejecta will be able to freely expand past the CSM. However, at regions of high CSM densities the ejecta will be decelerated by the circumstellar interaction. The result is that the interaction beneath the photosphere will continue to be the primary power source of the luminosity, but the spectral signatures of this interaction would be hidden until the photosphere has time to recede \citep{Andrews2018}. This may also explain the featureless blue continuum at early times that gives way to redshifted and broadened He features at later times as the photosphere recedes. The varying X-ray emission around 20 days past peak occurs at roughly the same time as the onset of these spectral features, which again may indicate that the photosphere has receded enough for X-rays generated by the CSM interaction to escape the ejecta \citep{Margutti2019}. It is interesting to note that at this phase \citet{Perley2019} estimate a blackbody radius of $\approx 18.5$ au, in very close agreement with the CSM inner radius of the model light curve from our grid that best matches the evolution of AT\,2018cow.

\subsection{Comparison with ZTF Rapidly Evolving Transient Sample}

\citet{Ho2021} constructed one of the largest samples of fast-evolving transients to date, with 22 spectroscopically classified objects in addition to 20 nonclassified ones. They found that fast-evolving transients can be split into three groups: faint and fast-evolving objects tend to be the initial shock-cooling phase seen in SNe IIb without the accompanying $^{56}$Ni-powered secondary peak, more luminous and slower-evolving objects tend to be interaction-powered SNe such as SNe Ibn and IIn, and the most luminous and fastest-evolving objects are radio-loud and X-ray luminous objects such as AT\,2018cow. In Figure \ref{fig:ztf} we compare our CSM model grid with the gold and silver samples from ZTF. Our models agree with their conclusions that the luminous and slower-evolving fast transients are dominated by interaction-powered SNe. However, we again show that our models reach even the most luminous and fast-evolving objects, including the parameter space of AT\,2018cow-like transients, implying a common origin between these objects and SNe Ibn.

\begin{figure*}
    \centering
    \includegraphics[width=0.7\textwidth]{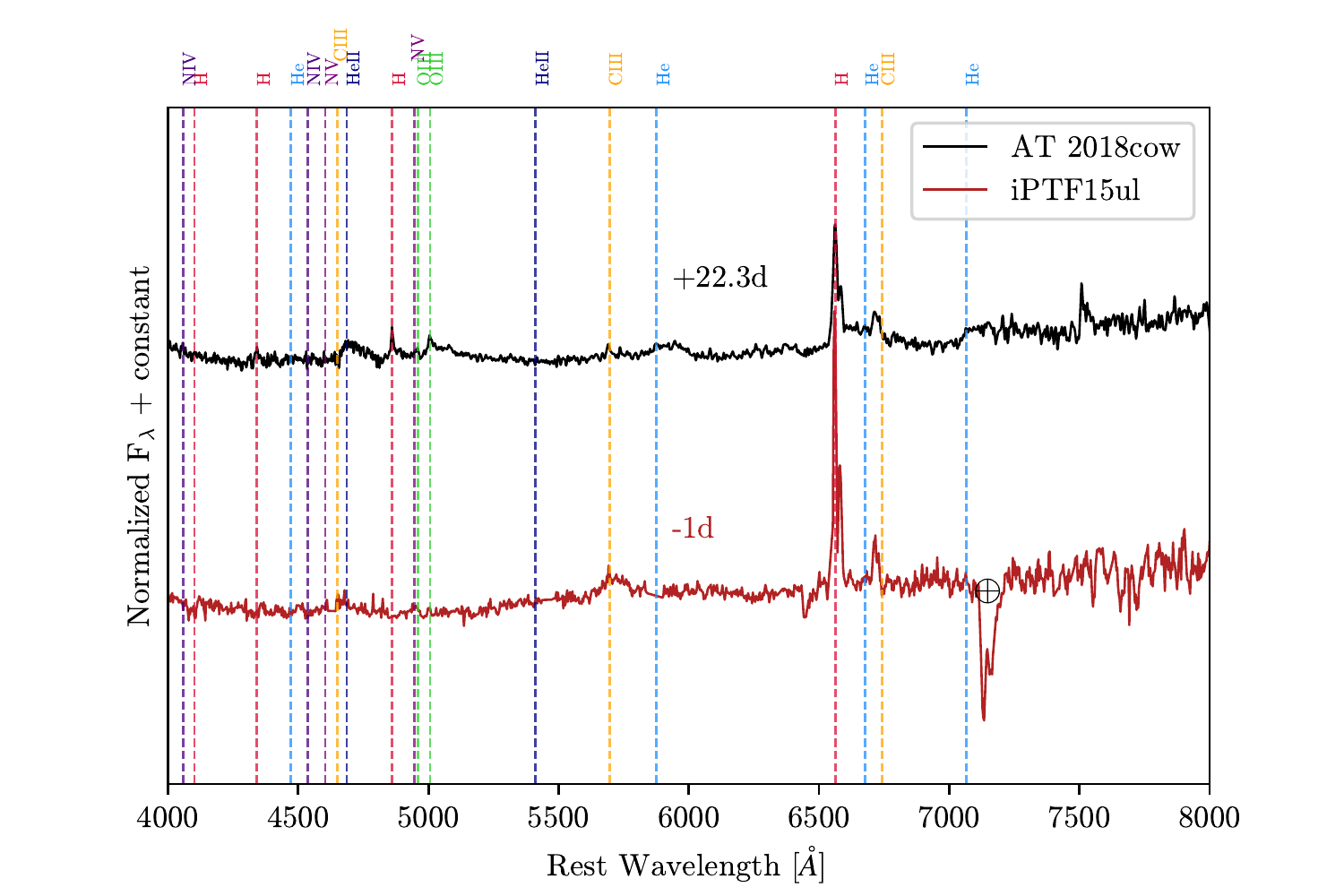}
    \caption{A dereddened rest-frame spectrum of the probable Type Ibn iPTF15ul around maximum light compared to a spectrum of AT\,2018cow about three weeks after maximum light. Phases relative to \textit{g}-band maximum light are denoted above each spectrum and some spectral features are marked with dashed colored lines. Telluric absorption is marked with a $\bigoplus$ symbol. Both objects display similar features, including highly ionized C and weak He lines, which set them apart from the other objects considered in this work. The spectrum of iPTF15ul was obtained from WISeREP \citep{Yaron2012}.}
    \label{fig:iptf15ul}
\end{figure*}

Several of the SNe we compare with our models in Figure \ref{fig:model_lcs} are included in either the ZTF spectroscopically classified sample or objects \citet{Ho2021} identify from literature as being fast transients. We have labeled these objects in Figure \ref{fig:ztf}. In this plot iPTF15ul stands out as being the fastest-evolving and most luminous spectroscopically classified object, besides AT\,2018cow. iPTF15ul was classified by \citet{Hosseinzadeh2017} as a probable SN Ibn and is one of the most luminous SNe Ibn reported to date. In Figure \ref{fig:iptf15ul} we compare a dereddened spectrum\footnote{Using the extinction law of \citet{Cardelli1989} and the estimated $A_{\text{V}}$ value from \citet{Hosseinzadeh2017}} of iPTF15ul around peak brightness to the spectrum of AT\,2018cow three weeks after peak. Similar features are seen between the two spectra, including C III emission lines and weak He I features, the latter of which is rare in SNe Ibn spectra at maximum light \citep{Hosseinzadeh2017,Gangopadhyay2020}. The spectra of these two objects appear more similar to each other than to any of the spectra shown in Figure \ref{fig:spec}.

Based on the similar light-curve properties and spectral features, iPTF15ul may be a transitional object between SNe Ibn and AT\,2018cow-like fast transients. However, several differences set iPTF15ul apart from AT\,2018cow-like objects. First, an X-ray search from the Swift X-ray Telescope \citep{Burrows2005} only yielded nondetections with an upper limit of 1.6 $\times$ 10$^{-2}$ counts s$^{-1}$ at maximum light. This may indicate some of the physical processes powering the high-energy emission seen in AT\,2018cow are missing in the case of iPTF15ul. Furthermore, its extremely high peak luminosity may be affected by host reddening estimates, which are highly uncertain \citep{Hosseinzadeh2017}. If the reddening estimate is correct, though, iPTF15ul shows that some SNe Ibn without luminous X-ray emission may still occupy the same region of parameter space as AT\,2018cow-like transients, even if the latter do have a distinct source of X-ray and radio emission. This is particularly important as future time-domain surveys will photometrically classify more objects across different regions of fast transient parameter space. Our model grid shows that these objects can be explained entirely by CSM interaction and radioactive decay on the basis of their light curves alone.

\subsection{Common Progenitor Scenarios}

The host galaxies of fast transients and SNe Ibn have been extensively studied \citep{Drout2014,Hosseinzadeh2019,Lyman2020,Wiseman2020,Ho2021}. The majority of spectroscopically classified fast transients from \citet{Ho2021} and the fast transients from D14 were found in star-forming galaxies. This indicates that the progenitors of most fast transients are massive stars. The range of parameters in both our model grid and the best-fit {\tt Minim} models can tell us more about the progenitor systems of these transients. We note a general trend in which fainter, slower-evolving, interaction-driven SNe have lower explosion energies (governed by $M_{\text{ej}}$ and $v_{\text{ej}}$) and less $^{56}$Ni produced. On the other hand, the models that best reproduce the observed behavior of the fastest and most luminous transients have fast ejecta, relatively low masses of both ejecta and CSM, and produce more $^{56}$Ni. The small ejecta mass ($\approx$1 $M_\odot$) perhaps could indicate much of the progenitor star's mass remains gravitationally bound to a compact remnant, as has been proposed for other fast-evolving transients \citep{Dexter2013}.

To gain a qualitative understanding of the proposed progenitor systems of fast transients and SNe Ibn, we compare our best-fit model parameters to the CSM properties inferred from observation. Signatures of mass loss in the months to years leading up to explosion have been observed for several SNe Ibn. In the first case, a preexplosion outburst was observed at the position of SN\,2006jc two years before explosion \citep{Foley2007,Pastorello2007,Smith2008}. However, the rise of SN\,2006jc was not well constrained, and from X-ray data we can infer that the shock did not reach the CSM until several weeks after explosion \citep{Immler2008}. SN\,2019uo also has precursor emission observed approximately a year before explosion \citep{Strotjohann2021}. The light curve from our model grid that best matches SN\,2019uo has an inner CSM radius $R_0$=65 au, similar to estimates derived from light-curve fits \citep{Gangopadhyay2020} and preexplosion mass loss \citep{Strotjohann2021}.

All of our models require a significant CSM mass relatively close to the progenitor star, indicating a large rate of mass loss shortly before explosion. An increase in mass-loss rates may be common in the years prior to interaction-driven SNe \citep{Ofek2014,Bruch2021,Strotjohann2021}, including eruptive mass-loss events \citep{Wang2020}. A proposed progenitor of SNe Ibn are WR stars \citep{Foley2007,Smith2012}, yet they have not been observed to undergo violent luminous blue variable-like eruptions. However, if such events occur during the nuclear burning stages within the last months to years of a WR star's lifetime \citep{Shiode2014}, they would not be observable in the Galactic WR population. On a qualitative level our model parameters agree with those from simulations of exploding WR stars, including low $M_{\text{ej}}$ and $M_{\text{Ni}}$ \citep{Dessart2011}. This may indicate that a WR-like progenitor to SNe Ibn and some fast transients is plausible.

Our model grid and the {\tt Minim} models predict different relationships between the properties of the CSM and the peak luminosities of the transients. This disagreement may be due to different assumptions made between the models: for instance, the different ejecta power-law indices or the use of a mass-loss rate to derive the CSM densities in the case of the {\tt Minim} models. For the model grid, the fact that more luminous models have smaller CSM radii may help to explain the peculiar features of AT\,2018cow-like transients, including the X-ray and radio emission and the delayed emergence of spectral features. These unusual features may be explained if the CSM is very close to the progenitor star and is quickly enveloped by the expanding ejecta. If this is the case, then the spectral features will not emerge until the photosphere recedes back past the CSM shell. This also explains the increased variability in the X-ray luminosity of AT\,2018cow beginning at the same phase, as once the photosphere recedes past the location of the shock within the CSM fewer X-rays are reprocessed by the ejecta. 

\section{Conclusions}\label{sec:conclusions}

We present one of the first investigations into a common powering mechanism between a sample of Type Ibn SNe and other photometrically classified fast optical transients. We are motivated to consider interaction with CSM as a powering mechanism for these two samples based on their similar light-curve properties and spectral features. We identify several fast-evolving Type Ibn SNe with well-sampled multiband light curves using data from LCO and Swift. We notice many similarities when comparing the light curves, colors, blackbody radii, and spectra of these Type Ibn SNe with those of fast transients such as AT\,2018cow. Modeling their light curves with luminosity inputs from circumstellar interaction and $^{56}$Ni decay reproduces the observed range of peak luminosities, rise times, and decline times of the objects in our sample, suggesting that these transients may have similar progenitor environments with significant mass-loss rates prior to explosion. 

These results are in agreement with recent studies \citep[e.g.,][]{Ho2021} which have found that fast transients are a heterogeneous class of objects, some of which show signatures of circumstellar interaction. Additionally, our models show that circumstellar interaction can reproduce the evolution of even the fastest-evolving and most luminous transients. The model parameters presented in this work demonstrate that relatively little ejecta mass and CSM ($\lesssim$4 $M_\odot$ total) are needed to reproduce the properties of fast transients, arguing against the need for exotic progenitor systems or powering sources to explain these objects. Additionally, models with faster-evolving light curves tend to have denser and more confined CSM, possibly indicating large-scale mass-loss events prior to explosion. However, it remains to be seen whether fast transients with luminous X-ray and radio emission, such as AT\,2018cow, can also be explained by circumstellar interaction, or if additional powering sources are needed to reproduce these features.

The analytical circumstellar-interaction models used in this work make several simplifying assumptions, such as a stationary photosphere and spherical symmetry, that are likely unrealistic. In the future, further work should be done in modeling circumstellar interaction with an asymmetric distribution of material, as this is both a more realistic physical scenario \citep{Smith2014} and will have important effects on observation \citep{Smith2017}. It is possible that an asymmetric CSM may be able to reproduce the full range of observed features of AT\,2018cow-like fast transients, but more work must be done to test this hypothesis.

This study demonstrates the importance of circumstellar interaction in understanding the properties of core-collapse SNe. It is likely that the majority of massive stars undergo enhanced mass loss at the ends of their lifetimes \citep{Ofek2014,Bruch2021,Strotjohann2021}, suggesting that circumstellar interaction in core-collapse SNe to some degree may be ubiquitous. This points to the growing need for more rapid spectroscopic follow-up of transients, especially fast-evolving objects at cosmological distances, in order to better understand the overlap between fast transients and interaction-powered classes of SNe.

\begin{acknowledgements}

We thank the anonymous reviewer for helpful comments and feedback. C.P., D.A.H., J.B., D.H, and E.P.G. are supported by NSF grants AST-1911225 and AST-1911151. I.A. is a CIFAR Azrieli Global Scholar in the Gravity and the Extreme Universe Program and acknowledges support from that program, from the European Research Council (ERC) under the European Union’s Horizon 2020 research and innovation program (grant agreement number 852097), from the Israel Science Foundation (grant number 2752/19), from the United States - Israel Binational Science Foundation (BSF), and from the Israeli Council for Higher Education Alon Fellowship. Research by S.V. is supported by NSF grants AST–1813176 and AST–2008108. This research makes use of observations from the Las Cumbres Observatory global telescope network as well as the NASA/IPAC Extragalactic Database (NED), which is operated by the Jet Propulsion Laboratory, California Institute of Technology, under contract with NASA.

\end{acknowledgements}

\textit{\software}{Astropy \citep{Astropy2013,Astropy2018}, \texttt{emcee} \citep{ForemanMackey2013}, \texttt{lcogtsnpipe} \citep{Valenti2016}, Matplotlib \citep{Hunter2007}, \texttt{Minim} \citep{Chatzopoulos2013}, NumPy \citep{Harris2020}, \texttt{SSS-CSM} \citep{Jiang2020}, Superbol \citep{Nicholl2018}}

\appendix 
\restartappendixnumbering

\begin{deluxetable}{lclllc}[h!]
\tablecaption{SNe Ibn Sample Parameters \label{tab:ibnsample}}
\tablehead{
\colhead{Object Name} & \colhead{Redshift} & \colhead{Time of Explosion} & \colhead{Time of Maximum} & \colhead{\textit{g}-band Decline Rate} & \colhead{Reference} \\ & & (MJD) & (MJD) & (mag day$^{-1}$) & }
\startdata
SN\,2019uo & 0.020 & 58499.4$\pm$1.39 & \tablenotemark{a}58508.1$\pm$0.5 & 0.126$\pm$0.005 & \citet{Gangopadhyay2020}\\
SN\,2019deh & 0.054 & 58579.99$\pm$0.25 & \tablenotemark{b}58588.5$\pm$0.65 & 0.112$\pm$0.003 & This work\\
SN\,2019wep & 0.025 & 58824.5$\pm$3.0& \tablenotemark{a}58828.5$\pm$2.0 & 0.145$\pm$0.003 & A. Gangopadhyay in prep. (2021)\\
SN\,2021jpk & 0.038 & 59316.3$\pm$0.99 & \tablenotemark{b}59324.10$\pm$0.45 & 0.147$\pm$0.002 & This work\\
\enddata
\tablenotetext{a}{\textit{r}-band}
\tablenotetext{b}{\textit{g}-band}
\end{deluxetable}

\begin{deluxetable}{lcccccccc}[h!]
\tablecaption{Optical Photometry \label{tab:optphot}}
\tablehead{
\colhead{Object Name} & \colhead{MJD} & \colhead{$U$} & \colhead{$B$} & \colhead{$g$} & \colhead{$V$} & \colhead{$r$} & \colhead{$i$} & \colhead{Telescope}}
\startdata
SN\,2019deh & 58587.6 & 16.56$\pm$0.06 & 17.37$\pm$0.1 & 17.36$\pm$0.01 & 17.49$\pm$0.03 & 17.6$\pm$0.01 & 17.75$\pm$0.02 & LCO \\
SN\,2019deh & 58587.6 & 16.6$\pm$0.06 & 17.61$\pm$0.03 & 17.37$\pm$0.01 & 17.49$\pm$0.03 & 17.6$\pm$0.01 & 17.75$\pm$0.02 & LCO \\
SN\,2019deh & 58588.9 & 16.66$\pm$0.06 & 17.66$\pm$0.03 & 17.32$\pm$0.01 & 17.5$\pm$0.04 & 17.57$\pm$0.02 & 17.65$\pm$0.03 & LCO \\
SN\,2019deh & 58588.9 & 16.72$\pm$0.06 & 17.65$\pm$0.03 & 17.33$\pm$0.01 & 17.63$\pm$0.04 & 17.46$\pm$0.02 & 17.71$\pm$0.04 & LCO \\
SN\,2019deh & 58590.5 & 16.88$\pm$0.07 & 17.86$\pm$0.04 & 17.59$\pm$0.02 & 17.7$\pm$0.04 & 17.74$\pm$0.03 & 17.8$\pm$0.04 & LCO \\
SN\,2019deh & 58590.5 & 16.76$\pm$0.07 & 17.87$\pm$0.04 & 17.64$\pm$0.02 & 17.63$\pm$0.04 & 17.72$\pm$0.03 & 17.8$\pm$0.04 & LCO \\
SN\,2019deh & 58593.5 & 17.07$\pm$0.07 & 18.3$\pm$0.05 & 17.97$\pm$0.03 & 18.04$\pm$0.06 & 18.06$\pm$0.05 & 18.11$\pm$0.06 & LCO \\
SN\,2019deh & 58593.5 & 17.11$\pm$0.09 & 18.25$\pm$0.05 & -- & 17.89$\pm$0.05 & 18.26$\pm$0.05 & 18.26$\pm$0.08 & LCO \\
SN\,2019deh & 58594.5 & 17.18$\pm$0.4 & -- & -- & -- & -- & -- & LCO \\
SN\,2019deh & 58599.9 & 18.03$\pm$0.08 & 18.83$\pm$0.05 & -- & -- & 19.09$\pm$0.04 & 19.13$\pm$0.06 & LCO \\
\enddata
\tablecomments{\textit{UBV} magnitudes are given in the Vega system while \textit{gri} magnitudes are given in the AB system. \\(This table is available in its entirety in machine-readable form.)}
\end{deluxetable}

\begin{deluxetable}{lcccc}[h!]
\tablecaption{Swift UV Photometry \label{tab:uvphot}}
\tablehead{
\colhead{Object Name} & \colhead{MJD} & \colhead{UVW2} & \colhead{UVM2} & \colhead{UVW1}}
\startdata
SN\,2021jpk & 59323.3 & 18.04$\pm$0.08 & 17.64$\pm$0.1 & --\\
SN\,2021jpk & 59332.5 & 19.11$\pm$0.13 & 19.19$\pm$0.16 & 19.16$\pm$0.13\\
SN\,2021jpk & 59335.9 & 19.2$\pm$0.29 & -- & 18.81$\pm$0.23\\
SN\,2021jpk & 59336.2 & 19.02$\pm$0.18 & 19.57$\pm$0.29 & 19.08$\pm$0.2\\
\enddata
\end{deluxetable}

\begin{deluxetable}{lccc}[h!]
\tablecaption{Log of FLOYDS Spectroscopic Observations \label{tab:spec}}
\tablehead{
\colhead{Object Name} & \colhead{MJD} & \colhead{Phase\tablenotemark{a}} & \colhead{Wavelength Range (\AA{})}}
\startdata
AT\,2018cow & 58309.33 & 22.3 & 3500--10,000 \\
SN\,2019uo & 58503.44 & -3.6 & 3500--9000 \\
SN\,2019uo & 58519.42 & 11.3 & 3500--10,000 \\
SN\,2019deh & 58587.53 & -1.0 & 3500--10,000 \\
SN\,2019deh & 58595.43 & 6.9 & 3500--10,000 \\
SN\,2019wep & 58826.49 & 0.0 & 3500--10,000 \\
SN\,2019wep & 58841.48 & 15.0 & 3500--10,000 \\
SN\,2021jpk & 59323.56 & 0.6 & 3500--10,000 \\
\enddata
\tablenotetext{a}{Days relative to \textit{g}-band maximum light}
\end{deluxetable}

\begin{deluxetable}{lcccccccc}[h!]
\tablecaption{Model Parameters\label{tab:modelparams}}
\tablehead{
\colhead{Model No.} & \colhead{$v_{\text{ej}}$ (km s$^{-1}$)} & \colhead{$M_{\text{ej}}$ ($M_\odot$)} & \colhead{$M_{\text{CSM}}$ ($M_\odot$)} & \colhead{$R_0$ (au)} & \colhead{log$_{10}$($\rho_0$) (g cm$^{-3}$)} & 
\colhead{$\epsilon$} & \colhead{$\kappa_\gamma$ (cm$^2$ g$^{-1}$)} & \colhead{$M_{\text{Ni}}$ ($M_\odot$)}}
\startdata
1 & 7680.00 & 2.00 & 0.70 & 41.00 & -11.66 & 0.01 & 0.13 & 0.03 \\
2 & 10,000.00 & 3.00 & 0.95 & 65.00 & -11.80 & 0.01 & 0.12 & 0.08 \\
3 & 10777.78 & 2.89 & 0.91 & 62.50 & -11.73 & 0.02 & 0.12 & 0.13 \\
4 & 11555.56 & 2.78 & 0.87 & 60.00 & -11.66 & 0.02 & 0.11 & 0.18 \\
5 & 12333.33 & 2.67 & 0.82 & 57.50 & -11.58 & 0.03 & 0.11 & 0.19 \\
6 & 13111.11 & 2.56 & 0.78 & 55.00 & -11.51 & 0.03 & 0.10 & 0.20 \\
7 & 13888.89 & 2.44 & 0.74 & 52.50 & -11.44 & 0.04 & 0.10 & 0.20 \\
8 & 14666.67 & 2.33 & 0.70 & 50.00 & -11.37 & 0.04 & 0.09 & 0.21 \\
9 & 15444.44 & 2.22 & 0.66 & 47.50 & -11.29 & 0.05 & 0.09 & 0.22 \\
10 & 16222.22 & 2.11 & 0.62 & 45.00 & -11.22 & 0.05 & 0.08 & 0.22 \\
11 & 17000.00 & 2.00 & 0.57 & 42.50 & -11.15 & 0.06 & 0.08 & 0.23 \\
12 & 17777.78 & 1.89 & 0.53 & 40.00 & -11.08 & 0.06 & 0.07 & 0.24 \\
13 & 18555.56 & 1.78 & 0.49 & 37.50 & -11.01 & 0.07 & 0.07 & 0.25 \\
14 & 19333.33 & 1.67 & 0.45 & 35.00 & -10.93 & 0.07 & 0.06 & 0.26 \\
15 & 20111.11 & 1.56 & 0.41 & 32.50 & -10.86 & 0.08 & 0.06 & 0.26 \\
16 & 20888.89 & 1.44 & 0.37 & 30.00 & -10.79 & 0.08 & 0.05 & 0.27 \\
17 & 21666.67 & 1.33 & 0.32 & 27.50 & -10.72 & 0.09 & 0.05 & 0.28 \\
18 & 22444.44 & 1.22 & 0.28 & 25.00 & -10.64 & 0.09 & 0.04 & 0.28 \\
19 & 23222.22 & 1.11 & 0.24 & 22.50 & -10.57 & 0.10 & 0.04 & 0.29 \\
20 & 24000.00 & 1.00 & 0.20 & 20.00 & -10.50 & 0.10 & 0.03 & 0.30 \\
\enddata
\end{deluxetable}
\pagebreak
\begin{deluxetable}{lccccccccc}
\tablecaption{Best-fit parameters for the {\tt hybrid} model computed with the {\tt Minim} code.}
\tablehead{
Object\tablenotemark{a} & $t_0$ & $R_{\rm ej}$ & $M_{\rm ej}$ & $M_{\rm csm}$ & $\dot{\rm M}$\tablenotemark{b} & $M_{\rm Ni}$ & $v_{\rm ej}$ & $\log_{10} \rho_{\rm CSM}$ & $R_{\rm out}$ \\
  & (day) & ($10^{13}$ cm) & ($M_\odot$) & ($M_\odot$) & ($M_\odot$yr$^{-1}$) & ($M_\odot$) & ($10^3$ km~s$^{-1}$) & (g~cm$^{-3}$) & ($10^{13}$ cm) \\
}
\startdata
AT\,2018cow & -0.06$\pm$0.03 & 64$\pm$12 & 0.67$\pm$0.03 & 0.68$\pm$0.03 & 0.99$\pm$0.26 & 0.036$\pm$0.001 & 56.5$\pm$1.6 & -10.81$\pm$0.21 & 70$\pm$13 \\
iPTF15ul & 0.00$\pm$0.12 & 95$\pm$9 & 1.47$\pm$0.08 & 0.45$\pm$0.040 & 1.80$\pm$0.11 & 0.190$\pm$0.012 & 69.3$\pm$4.6 & -10.99$\pm$0.11 & 96$\pm$9 \\
LSQ13ddu & -0.90$\pm$0.27 & 41$\pm$4 & 0.89$\pm$0.06 & 0.49$\pm$0.05 & 1.27$\pm$0.17 & 0.034$\pm$0.001 & 33.8$\pm$1.0 & -10.41$\pm$0.15 & 42$\pm$4 \\
SN\,2019deh & -0.96$\pm$0.13 & 76$\pm$3 & 1.03$\pm$0.03 & 0.98$\pm$0.04 & 1.97$\pm$0.14 & 0.031$\pm$0.001 & 36.1$\pm$0.5 & -10.76$\pm$0.07 & 77$\pm$3 \\
SN\,2019uo & 0.82$\pm$0.17 & 22$\pm$5 & 1.37$\pm$0.26 & 0.28$\pm$0.06 & 1.70$\pm$0.35 & 0.005$\pm$0.001 & 17.4$\pm$1.0 & -9.76$\pm$0.30 & 23$\pm$5 \\
SN\,2019wep & 1.57$\pm$0.59 & 19$\pm$9 & 1.38$\pm$0.36 & 0.21$\pm$0.13 & 4.18$\pm$2.77 & 0.015$\pm$0.005 & 20.1$\pm$1.9 & -9.22$\pm$0.85 & 19$\pm$9 \\
SN\,2021jpk & -0.96$\pm$0.04 & 40$\pm$2 & 1.02$\pm$0.08 & 0.47$\pm$0.01 & 1.11$\pm$0.29 & 0.000$\pm$0.001 & 15.2$\pm$0.6 & -10.46$\pm$0.14 & 41$\pm$2 \\
\enddata
\label{tab:minimfit}
\tablenotetext{a}{$t_0$: time-shift; $R_{\rm ej}$: initial ejecta radius; $M_{\rm ej}$: ejecta mass; $M_{\rm csm}$: CSM mass; $\dot{\rm M}$: mass-loss parameter; \\
$M_{\rm Ni}$: $^{56}$Ni mass; $v_{\rm ej}$: ejecta velocity; $\log_{10}\rho_{\rm CSM}$: CSM density; $R_{\rm out}$: outer CSM radius}
\tablenotetext{b}{Used as a parameter for the CSM density assuming $\dot{{\rm M}} = 4 \pi {\rm R}^2_{\rm ej} \rho_{\rm CSM} v_{\rm w}$ and $v_w = 10$ km~s$^{-1}$}
\end{deluxetable}
\pagebreak
\section{\texttt{Minim} Model Light Curves}
\begin{figure*}
\gridline{\fig{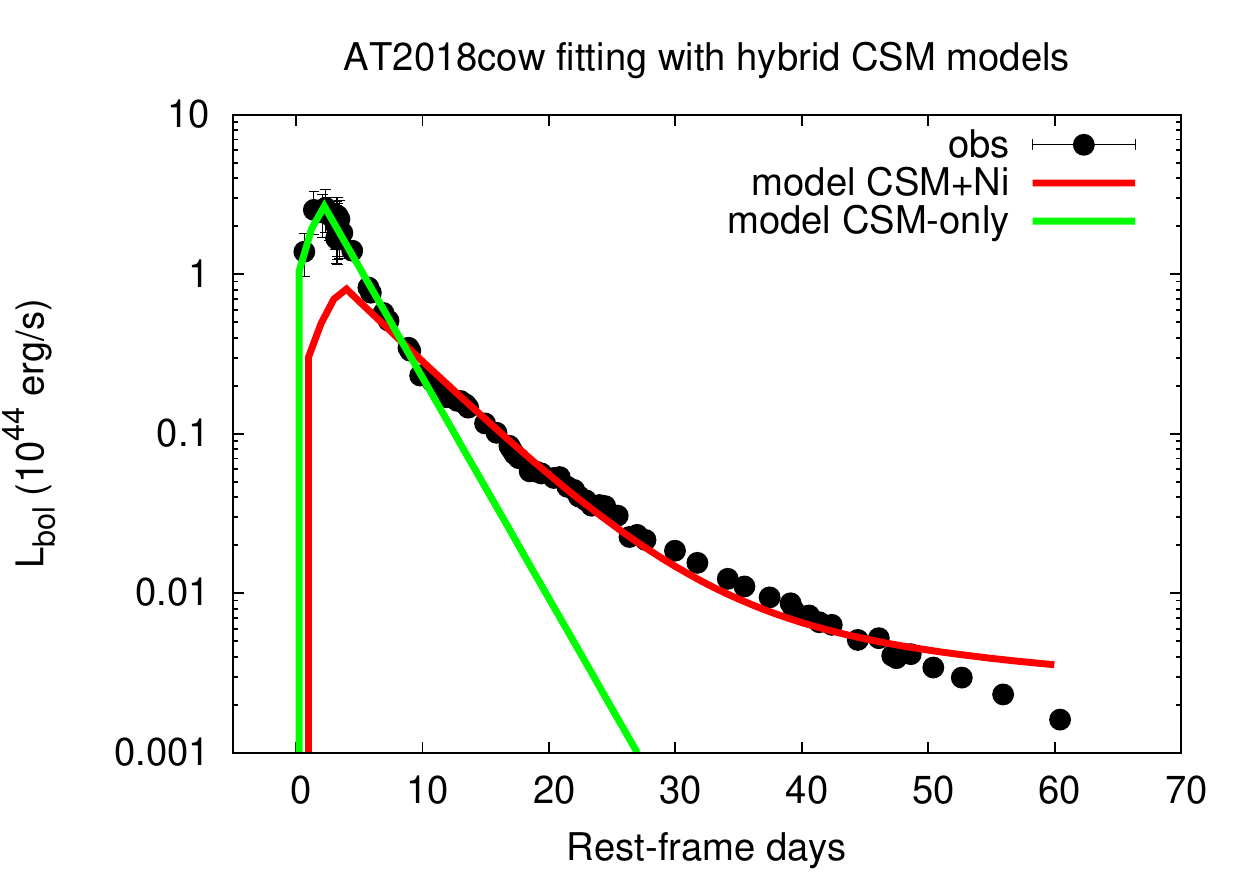}{0.3\textwidth}{(a)}
          \fig{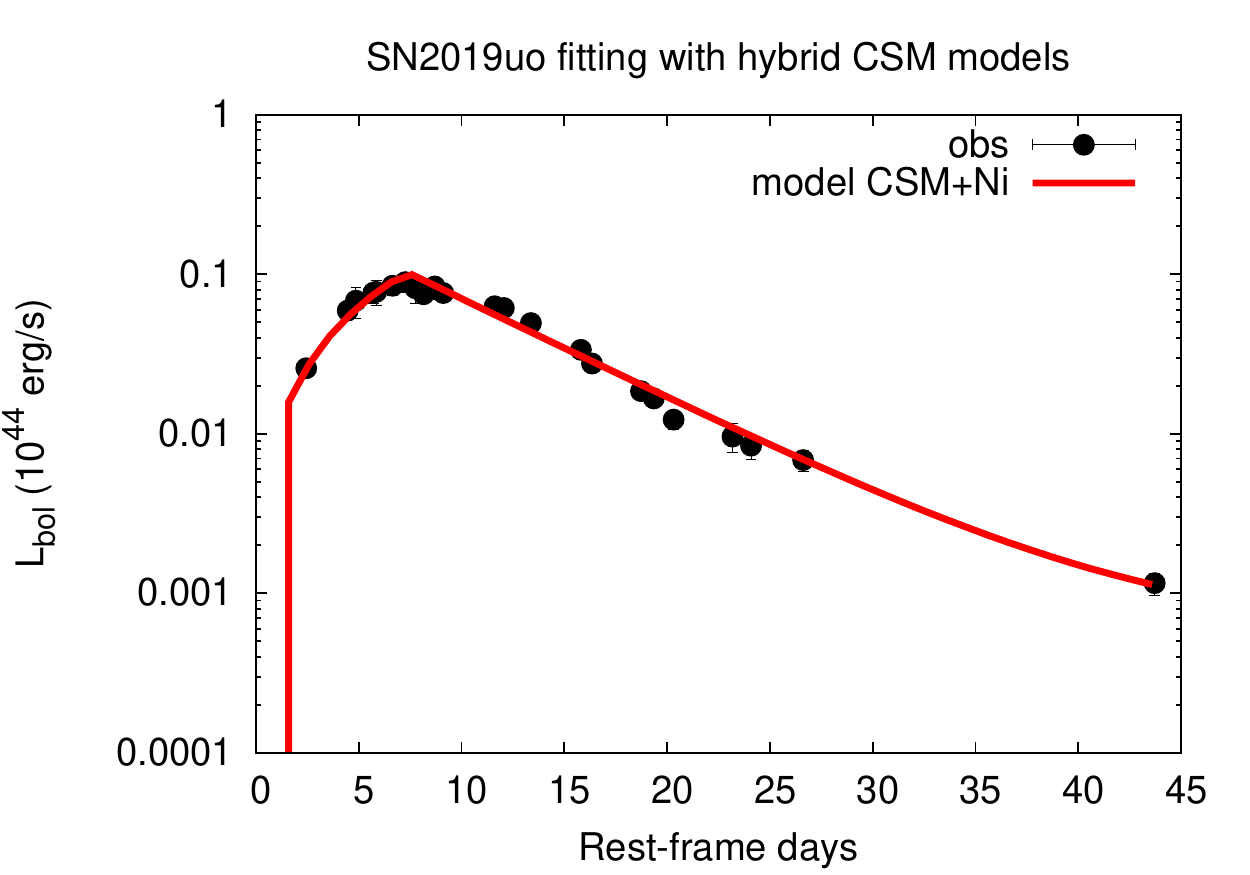}{0.3\textwidth}{(b)}
          \fig{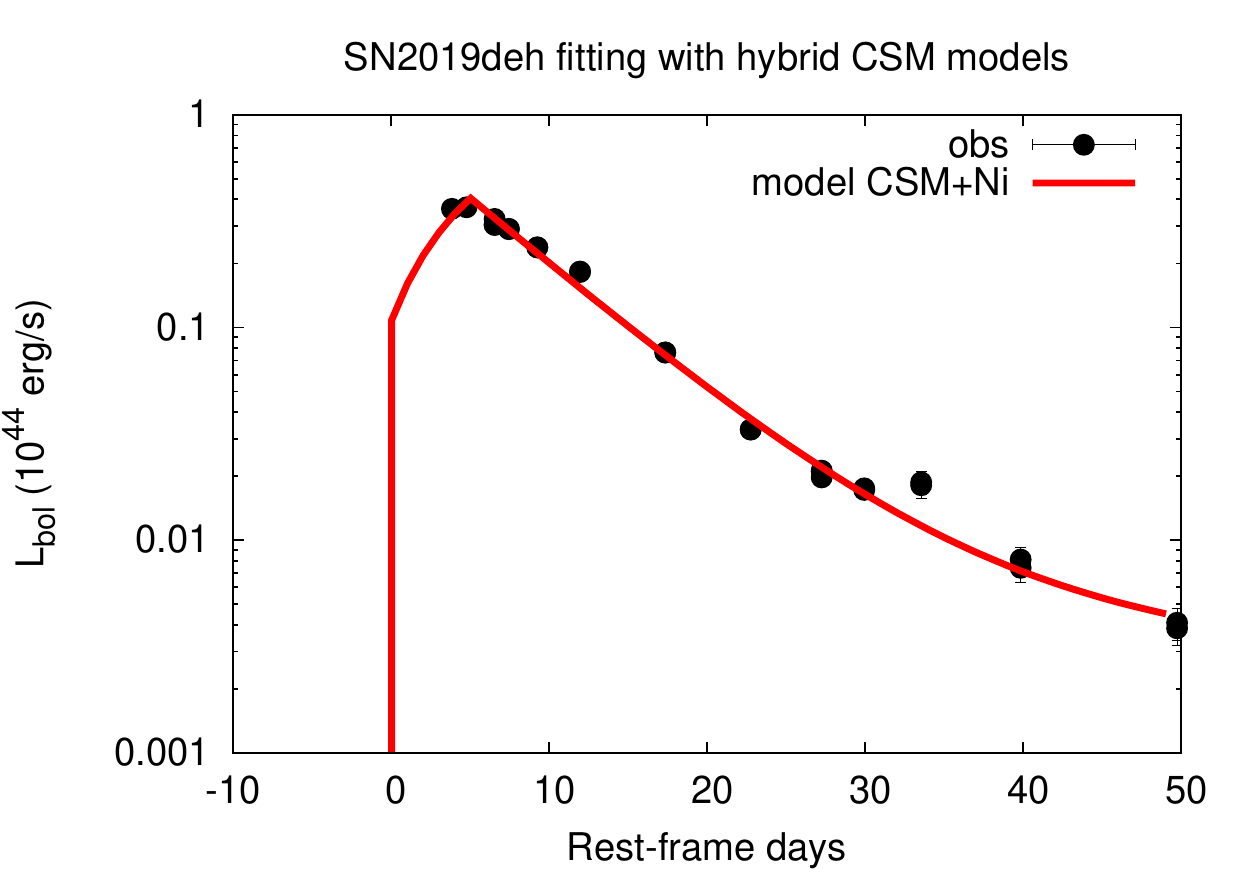}{0.3\textwidth}{(c)}
          }
\gridline{\fig{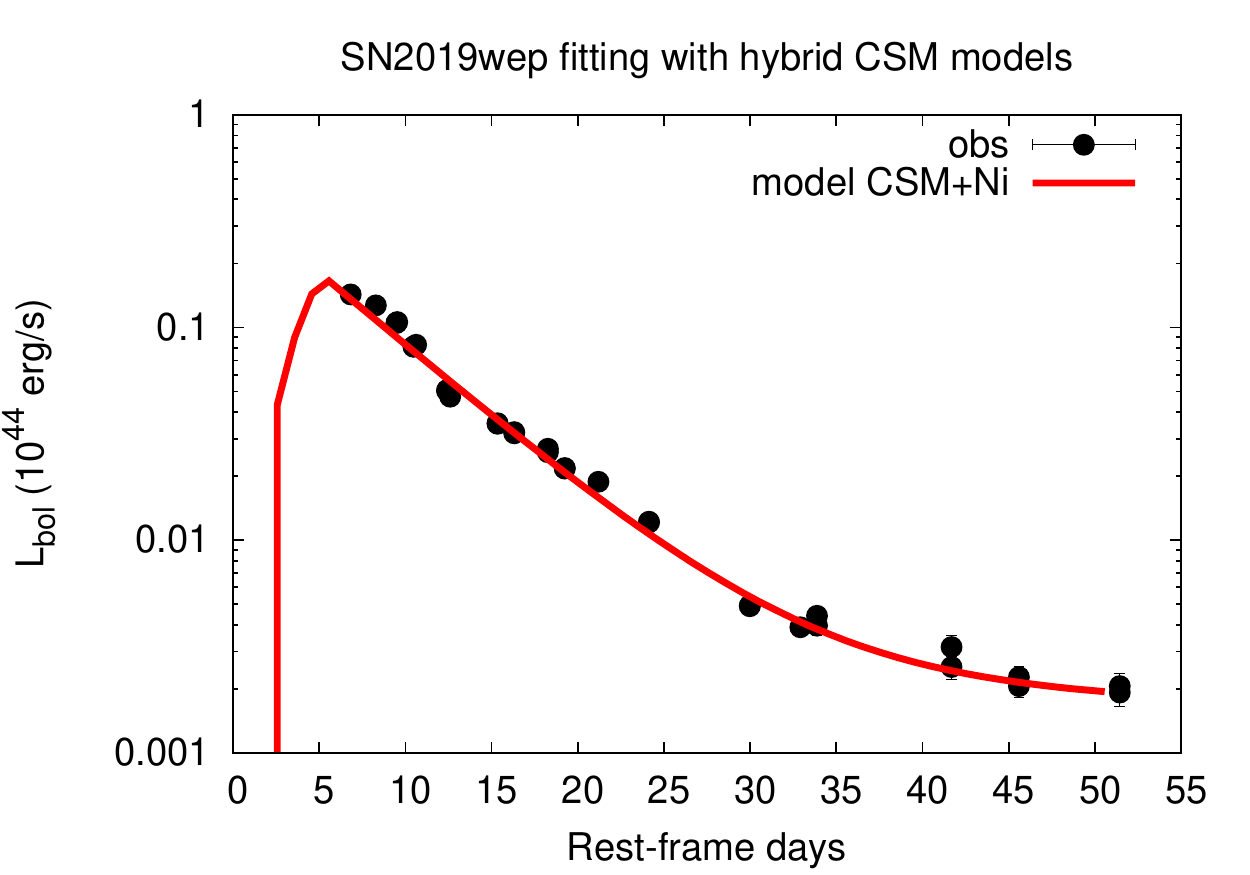}{0.3\textwidth}{(d)}
          \fig{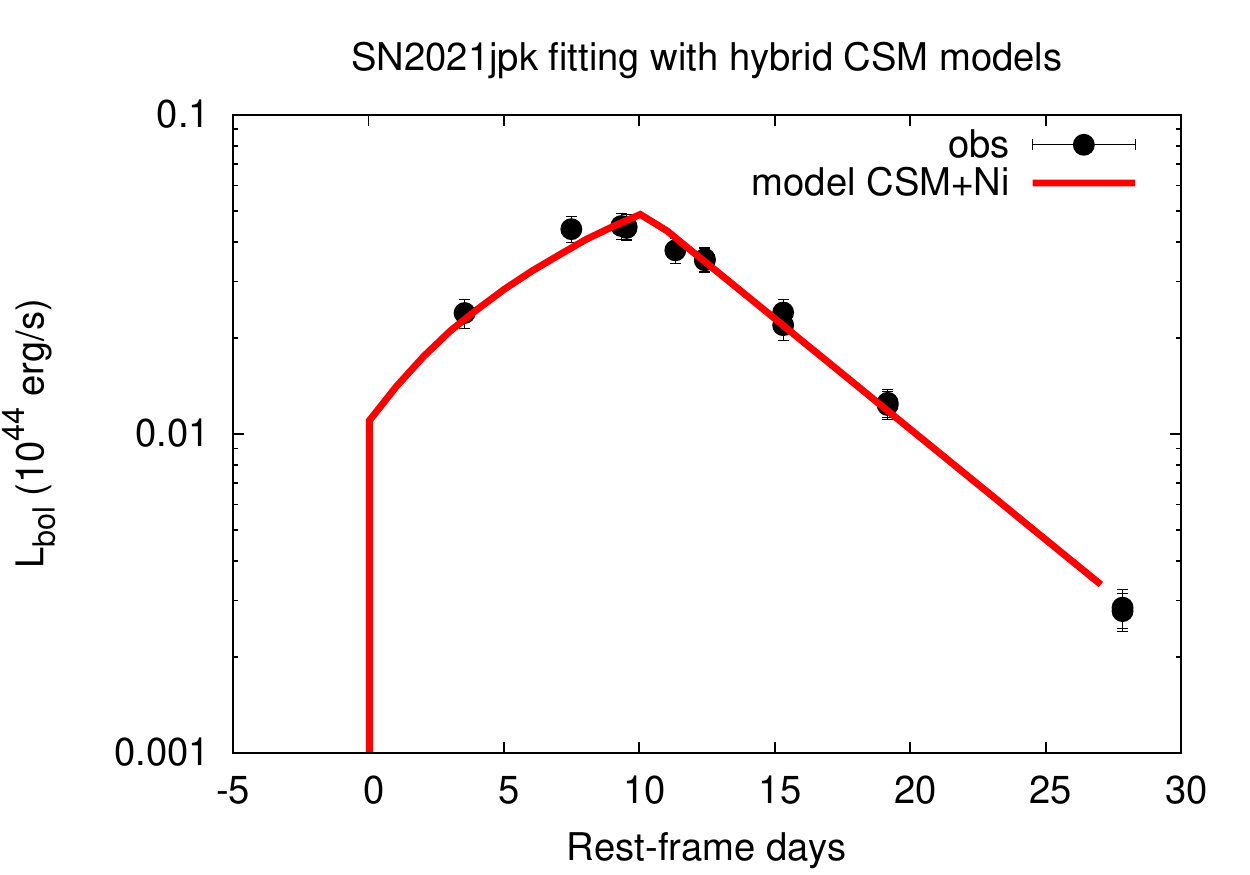}{0.3\textwidth}{(e)}
          \fig{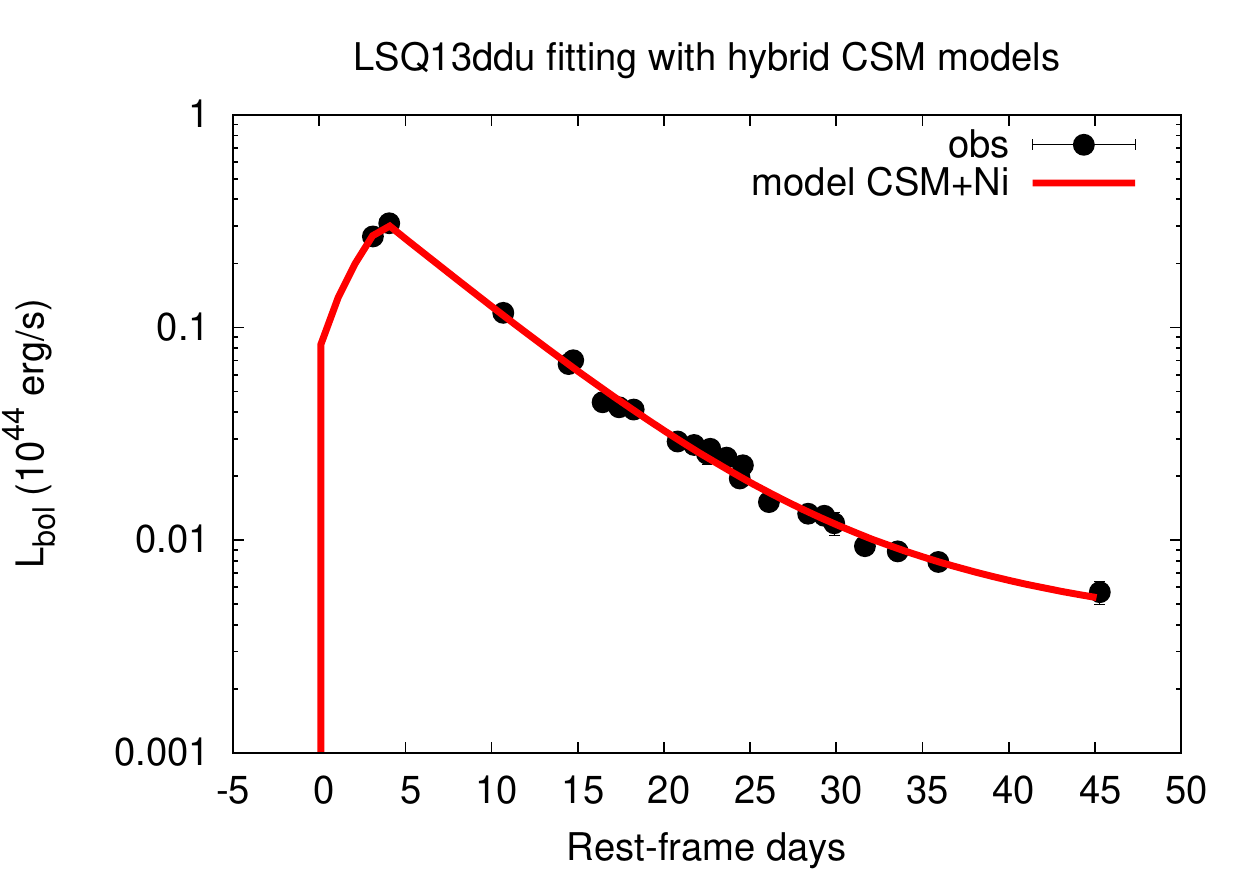}{0.3\textwidth}{(f)}
          }
\gridline{\fig{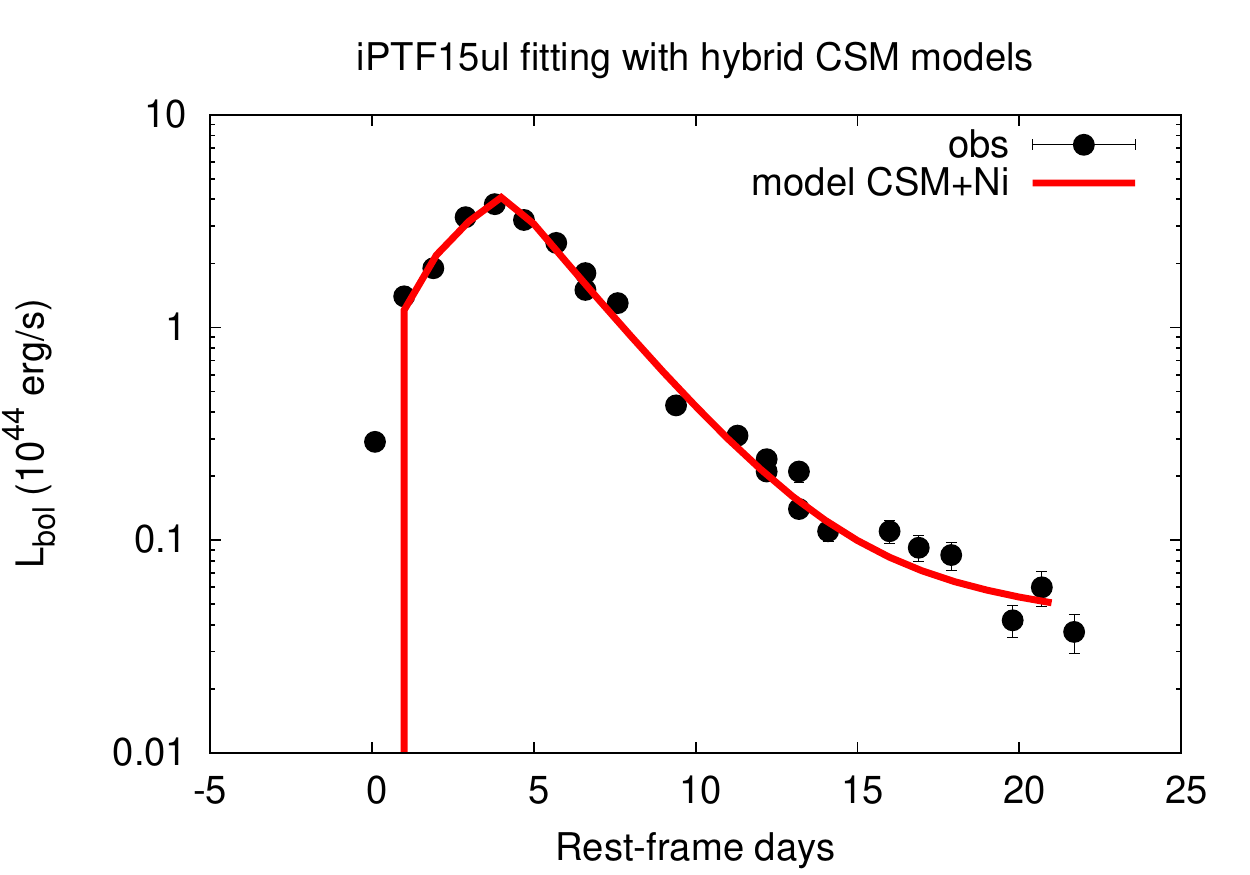}{0.3\textwidth}{(g)}}
\caption{Best-fit \texttt{Minim} models to the bolometric light curves of (a) AT\,2018cow, (b) SN\,2019uo, (c) SN\,2019deh, (d) SN\,2019wep, (e) SN\,2021jpk, (f) LSQ13ddu, and (g) iPTF15ul assuming luminosity contributions from CSM interaction and $^{56}$Ni decay.
\label{fig:minimfits}}
\end{figure*}

Here we show our best-fit \texttt{Minim} models in Figure \ref{fig:minimfits}.
\pagebreak

\end{document}